\documentclass[preprint,showpacs,preprintnumbers,superscriptaddress,amsmath,amssymb]{revtex4}

\usepackage{epsfig,amsfonts,amsmath}
\usepackage{dcolumn}
\usepackage{bm}

\begin{document}


\title{Spin Needlets for Cosmic Microwave Background Polarization Data Analysis}

\author{Daryl Geller}
\email{daryl@math.sunysb.edu}
\affiliation{Department of Mathematics, Stony Brook University, Stony Brook, NY 11794-3651, USA}
\author{Frode K. Hansen}
\email{frodekh@astro.uio.no}
\affiliation{Institute of Theoretical Astrophysics, University of Oslo, PO Box 1029 Blindern, N-0315 Oslo, Norway}
\author{Domenico Marinucci}
\email{marinucc@mat.uniroma2.it}
\affiliation{Dipartimento di Matematica, Universit\`a di Roma `Tor Vergata', Via della Ricerca Scientifica 1, I-00133 Roma, Italy}
\author{Gerard Kerkyacharian}
\email{kerk@math.jussieu.fr} \affiliation{Laboratoire de
Probabilit\`e et Mod\`eles Al\`eatoires}
\author{Dominique Picard}
\email{picard@math.jussieu.fr} \affiliation{Universit\`e Paris 7 and
Laboratoire de Probabilit\`e et Mod\`eles Al\`eatoires}

\date{\today}

\begin{abstract}
Scalar wavelets have been used extensively in the analysis of Cosmic
Microwave Background (CMB) temperature maps. Spin needlets are a new
form of (spin) wavelets which were introduced in the mathematical
literature by Geller and Marinucci (2008) as a tool for the analysis
of spin random fields. Here we adopt the spin needlet approach for
the analysis of CMB polarization measurements. The outcome of
experiments measuring the polarization of the CMB are maps of the
Stokes Q and U parameters which are spin 2 quantities. Here we
discuss how to transform these spin 2 maps into spin 2 needlet
coefficients and outline briefly how these coefficients can be used
in the analysis of CMB polarization data. We review the most
important properties of spin needlets, such as localization in pixel
and harmonic space and asymptotic uncorrelation. We discuss several
statistical applications, including the relation of angular power
spectra to the needlet coefficients, testing for non-Gaussianity on
polarization data, and reconstruction of the $E$ and $B$ scalar
maps.
\end{abstract}

\maketitle

\setcounter{MaxMatrixCols}{10}

\section{Introduction}

The WMAP satellite has provided the scientific community with the highest
resolution full-sky data of the Cosmic Microwave Background (CMB) obtained
to date (\cite{hinshaw}). These data have allowed precise estimates of the
temperature angular power spectrum $C_{\ell }$ up to the third Doppler peak
and thereby high precision measurements of many cosmological parameters. In
addition to measuring the temperature fluctuations in the CMB with high
sensitivity, WMAP has also measured the polarization of the background
radiation. But the polarization signal is almost an order of magnitude
smaller than the temperature signal and is therefore measured with much
lower sensitivity. The low signal-to-noise level makes it important to have
a good understanding of systematic effects, in particular correlated noise
properties. Little is so far known about polarized foregrounds. For low
signal-to-noise data, small errors in the understanding of systematic
effects and foregrounds could lead to errors in the estimates of the
polarization angular power spectra.

In the near future, the situation with respect to CMB polarization will
improve significantly. The Planck satellite will take full-sky measurements
of the polarized CMB sky (\cite{laureijs}) and several other ground/balloon
based experiments will follow up with high sensitivity observations in
smaller regions of the sky (the CLOVER, QUIET and QUAD experiments to
mention a few). Both ESA and NASA are planning high sensitivity full-sky
satellite borne experiments within the next 10-20 years.

Polarization measurements allow for estimation of three more angular power
spectra, the $C_{\ell }^{TE}$ power spectrum measuring the correlation
between temperature and $E$ mode polarization as well as $C_{\ell }^{EE}$
and $C_{\ell }^{BB}$ which are the spectra of the $E$ and $B$ modes of
polarization. The $TE$ and $EE$ spectra at large scales are used to estimate
the reionization optical depth to high precision, a parameter to which the
temperature power spectrum is not very sensitive. These spectra also give
independent measurements of the other cosmological parameters estimated from
the temperature power spectrum. This does not only serve as a consistency
check, but also improves the statistical error bars on these parameters.
Finally the much weaker $BB$ polarization power spectrum may at large scales
be dominated by a signal arising from a background of gravitational waves
which originated from inflation. This would be an important confirmation of
the theory of inflation, and would as well give information about the detailed
physics of the inflationary epoch. At smaller scales, the $BB$ spectrum is
dominated by $EE$ modes which have been converted to $BB$ modes by
gravitational lensing of large scale structure in the more recent universe.

Taking into account the huge amount of polarization data which will be
available in the next 1-2 decades, as well as the important cosmological
information contained in these data, it is clear that efficient data
analysis tools will be necessary. While a large amount of data analysis
techniques have been developed for analyzing CMB temperature data, much less
attention has been given to analysis techniques for polarization data, also
due to the lack of available high sensitivity observations.

An important tool for the analysis of temperature data has been various
kinds of spherical wavelets (see \cite{antoine,Antodem,jfaa1} and the
references therein). Indeed, in the last decade, wavelet methods have found
applications in virtually all areas where statistical methods for CMB data
analysis are required. Just to mention a few examples, we recall foreground
subtraction (\cite{gorski}), point source detection (\cite{vielva03,sanz}),
testing for non-Gaussianity (see \cite{vielva04,chmpv,jjin}), search for
anisotropies \cite{Cruz1,Cruz2}, component separation (\cite{moudden}),
cross-correlation between CMB and Large Scale Structure Data (\cite{mcewen1}%
), and many others. Directional wavelets have been advocated by \cite%
{mcewen2,wiaux2}. The rationale for such a widespread interest can be
explained as follows: CMB models are best analyzed in the frequency domain,
where the behaviour at different multipoles can be investigated separately;
on the other hand, partial sky coverage and other missing observations make
the evaluation of exact spherical harmonic transforms troublesome. The
combination of these two features makes the time-frequency localization
properties of wavelets most valuable.

More recently, a new kind of wavelets has found many fruitful
applications in the CMB literature, the so-called spherical
needlets. Spherical needlets were introduced in the functional
analysis literature by \cite{npw1,npw2}, and then considered for the
statistical analysis of spherical random fields by
(\cite{bkmpAoS,bkmpBer}), with a view to applications to the
statistical analysis of CMB\ data, including estimation of the
angular power spectrum, testing for Gaussianity and bootstrap
procedures. The related literature is already quite rich:
applications to the analysis of the integrated Sachs-Wolfe effect by
means of cross-correlation between CMB\ and Large Scale Structure
data were given in \cite{pbm06}; a general introduction to their use
for CMB data analysis is given in \cite{mpbb08}; a discussion on
optimal weight functions is provided in \cite{guilloux}; further
applications include (\cite{lan}, \cite{dela08,fay08,fg08}). The
extension to the construction of \cite{npw1,npw2} to the case of
non-compactly supported weight functions is provided by
\cite{gmcw,gm2,gm1}, where the relationship with Spherical Mexican
Hat Wavelets is also investigated, leading to the analysis of
so-called \emph{Mexican Needlets.} The stochastic properties of the
corresponding Mexican needlet coefficients are established in
\cite{lan2,Mayeli}.

In this paper, we show how the scalar needlets which were applied to CMB
temperature data can be extended to polarization using \textit{spin
needlets. }The latter were introduced in the mathematical literature by \cite%
{spin-mat}\textit{; }they can be viewed as spin-2 wavelets designed for
spin-2 fields on the sphere having the same properties when applied to CMB
polarization fields as scalar needlets have when applied to CMB temperature
fields. In particular, we shall show below that spin needlets enjoy both the
localization and the uncorrelation properties that make scalar needlets a
powerful tool for the analysis of CMB temperature data. The aim of this
paper is then to introduce the spin-2 needlets, show how they can be applied
to \emph{Q} and \emph{U} maps, and discuss some preliminary ideas for
possible future statistical applications (how these maps can be
reconstructed from the spin-2 needlet coefficients, how \emph{EE}
and \emph{BB} spectra may be obtained directly from these coefficients,
tests of non-Gaussianity and others).

The plan of the paper is as follows: in Section 2 we provide a quick review
of standard (scalar) needlets and their main properties; in Section 3 we
report the main results of \cite{spin-mat}, where spin needlets were first
introduced and investigated from the mathematical point of view. In Section
4 we discuss a number of possible future applications to polarization data
analysis; in Section 5 we provide a comparison with alternative approaches
for the wavelet analysis of polarization data, while in Section 6 we
present some preliminary evidence on the reconstruction properties of spin
needlets. Some background material on spin spherical harmonics is given in
the Appendix.

\section{A review of scalar needlets}

To ease comparisons with the spin needlets which we will describe later, we
recall very briefly the construction of a needlet basis. The spherical
needlet (function) is defined as
\begin{equation}
\psi _{jk}(\hat{\gamma})=\sqrt{\lambda _{jk}}\sum_{\ell }b(\frac{\ell }{B^{j}%
})\sum_{m=-\ell }^{\ell }\overline{Y}_{\ell m}(\hat{\gamma})Y_{\ell m}(\xi
_{jk})\text{ };  \label{needlets_expansion}
\end{equation}%
here, $\hat{\gamma}$ is a direction $(\theta ,\phi )$ on the sphere, and $j$
is the frequency (multipole range) of the needlet. We use $\left\{ \xi
_{jk}\right\} $ to denote a set of \emph{cubature points }on the sphere,
corresponding to a given frequency $j$. In practice, we will identify these
points with the pixel centres in the HEALPix pixelization scheme \cite{healpix}. The
cubature weights $\lambda _{jk}$ are inversely proportional to the number of
pixels we are actually considering (see \cite{pbm06}, \cite{bkmpBer} for
more details); we can take for simplicity $\lambda _{jk}=4\pi /N_{j},$ where
here and throughout the paper we shall take $N_{j}$ to be the number of
cubature points $\left\{ \xi _{jk}\right\} .$ Needlets can then be viewed as
a combination, analogous to
convolution, of the projection operators $\sum_{m=-\ell }^{\ell }\overline{Y}%
_{\ell m}(\hat{\gamma})Y_{\ell m}(\xi _{jk})$ with a suitably chosen window
function $b(x)$ with $x=\ell/B^j$. The number $B$ appearing in the argument of the function $%
b(x)$ is a parameter which defines the needlet basis, as will be
discussed below. Special properties of $b(x)$ ensure that the
needlets enjoy quasi-exponential localization properties in pixel
space. Formally, we must ensure that (\cite{npw1,npw2}):

\begin{itemize}
\item \textbf{A1} The function $b^{2}(x)$ is positive in the range $x=[\frac{%
1}{B},B],$ zero otherwise; hence $b(\frac{\ell }{B^{j}})$ is positive in $%
\ell \in \lbrack B^{j-1},B^{j+1}]$

\item \textbf{A2 }the function $b(x)$ is infinitely differentiable in $%
(0,\infty ).$

\item \textbf{A3 }we have%
\begin{equation}
\sum_{j=1}^{\infty }b^{2}(\frac{\ell }{B^{j}})\equiv 1\text{ for all }\ell
>B.  \label{partun}
\end{equation}
\end{itemize}

Condition A1 can be generalized to cover functions such as $x\exp (-x)$, thus
leading to so-called Mexican needlets, see \cite{gmcw,gm2,gm1} where advantages
and disadvantages of such choice are also discussed. In the present
formulation, A1 ensures the needlets have bounded support in the harmonic
domain; A2 is needed for the derivation of the localization properties,
which we shall illustrate in the following section. Finally, A3 (\emph{the
partition of unity property}) is needed to establish the reconstruction
formula (\ref{recfor}). Examples of constructions satisfying A1-A3 are given
in \cite{pbm06,guilloux}.

We shall now recall briefly some of the general features of needlets, which
are not in general granted by other spherical wavelet constructions. We
refer for instance to \cite{mpbb08} for more details and a general
introduction for a CMB readership. Briefly, we recall the following features:

a) needlets do not rely on any tangent plane approximation (compare \cite%
{sanz}), and take advantage of the manifold structure of the sphere;

b) being defined in harmonic space, they are computationally very
convenient, and inherently adapted to standard packages such as HEALPix;

c) they allow for a simple reconstruction formula (see (\ref{recfor})),
where the same needlet functions appear both in the direct and the inverse
transform. This property is the same as for spherical harmonics but it is
\emph{not }shared by other wavelets systems;

d) they are quasi-exponentially (i.e. faster than any polynomial)
concentrated in pixel space, see (\ref{expine}) below;

e) they are exactly localized on a finite number of multipoles; the width of
this support is explicitly known and can be specified as an input parameter
(see (\ref{needlets_expansion}));

f) random needlet coefficients can be shown to be asymptotically
uncorrelated (and hence, in the Gaussian case, independent) at any
fixed angular distance, when the frequency increases (see
\cite{bkmpAoS}). This capital property can be exploited in several
statistical procedures, as it allows one to treat needlet
coefficients as a sample of independent and identically distributed
coefficients on small scales, at least under the Gaussianity
assumption.

More precisely, random needlet coefficients are given by
\begin{eqnarray}
\beta _{jk} &=&\int_{S^{2}}T(\hat{\gamma})\psi _{jk}(\hat{\gamma})d\hat{%
\gamma}  \label{needproj} \\
&=&\sqrt{\lambda _{jk}}\sum_{\ell }b(\frac{\ell }{B^{j}})\sum_{m=-\ell
}^{\ell }\left\{ \int_{S^{2}}T(\hat{\gamma})\overline{Y}_{\ell m}(\hat{\gamma%
})d\hat{\gamma}\right\} Y_{\ell m}(\xi _{jk})  \notag \\
&=&\sqrt{\lambda _{jk}}\sum_{\ell }b(\frac{\ell }{B^{j}})\sum_{m=-\ell
}^{\ell }a_{\ell m}Y_{\ell m}(\xi _{jk}).  \label{needcof}
\end{eqnarray}%
Here $j$ denotes the frequency of the coefficient and $k$ refers to
the direction $(\theta ,\phi )$ on the sky. The index $k$ can in
practice be the pixel number on the HEALPix grid. It is very
important to stress that, although the needlets do \emph{not} make
up an orthonormal basis for square
integrable functions on the sphere, they do represent a \emph{tight frame. }%
In general, a tight frame on the sphere is a countable set of functions
which preserves the norm; frames do not in general make up a basis, as they
admit redundant elements. They can be viewed as the closest system to a
basis, for a given redundancy, see \cite{hw96,bkmpAoS, gm1} for further
definitions and discussion. In our framework, the norm-preserving property
becomes
\begin{equation}
\sum_{jk}\beta _{jk}^{2}\equiv \int_{S^{2}}T^{2}(\hat{\gamma})d\hat{\gamma}%
=\sum_{\ell =1}^{\infty }(2\ell +1)\widehat{C}_{\ell }\text{ ,}
\label{frame}
\end{equation}%
where
\begin{equation*}
\widehat{C}_{\ell }=\frac{1}{2\ell +1}\sum_{m}|a_{\ell m}|^{2}
\end{equation*}%
is the raw angular power spectrum of the map $T(\hat{\gamma})$. (\ref%
{frame}) suggests immediately some procedures for angular power spectrum
estimations and testing (\cite{bkmpAoS,fg08}) and is related to a much more
fundamental result, i.e. the reconstruction formula
\begin{equation}
T(\hat{\gamma})\equiv \sum_{j,k}\beta _{jk}\psi _{jk}(\hat{\gamma})
\label{recfor}
\end{equation}%
which in turn is a non-trivial consequence of the careful construction
leading to (\ref{partun}). Again, we stress that the simple reconstruction
formula of (\ref{recfor}) is typical of tight frames but does not hold in
general for other wavelet systems. It is easy to envisage many possible
applications of (\ref{recfor}) when handling masked data.

\subsection{Localization and uncorrelation properties}

The following quasi-exponential localization property of needlets is due to %
\cite{npw1,npw2} and motivates their name:

For any $M=1,2,...$ there exists a positive constant $c_{M}$ such that for
any point $\hat{\gamma}\in S^{2}$ we have%
\begin{equation}
|\psi _{jk}(\hat{\gamma})|\leq \frac{c_{M}B^{j}}{(1+B^{j}\arccos (|\hat{%
\gamma}-\xi _{jk}|))^{M}}\text{ .}  \label{expine}
\end{equation}

We recall that $\arccos (|\hat{\gamma}-\xi _{jk}|)$ is just the geodesic
distance on the unit sphere between the position $\hat{\gamma}$ and the
position $\xi _{jk};$ (we recall in practice $\xi _{jk}$ can be the pixel
center of a HEALPix pixel $k$). (\ref{expine}) is then stating that, for any
fixed nonzero geodesic distance, the value of $\psi _{jk}(\hat{\gamma})$ goes to zero
faster than any polynomial (\emph{quasi-exponentially}) in the parameter $B.$
Thus needlets achieve excellent localization properties in both the real and
the harmonic domain. In \cite{gmcw}, (\ref{expine}) is extended to the case
of a non-compactly supported but smooth $b(x),$ thus covering also the
Mexican needlet case (where $b(x)\simeq x\exp (-x)).$

From the stochastic point of view, the crucial uncorrelation
properties for random spherical needlet coefficients were given in
\cite{bkmpAoS}. More precisely, two forms of uncorrelation were
established

\begin{itemize}
\item \textbf{P1} Whenever $\left| j_{1}-j_{2}\right| \geq 2,$ we have that $%
\left\langle \beta _{j_{1}k_{1}}\beta _{j_{2}k_{2}}\right\rangle =0$, $%
\left\langle {}\right\rangle $ denoting the expected value

\item \textbf{P2} For $j_{1}=j_{2},$ for any $M>0$ there exist a constant $%
C_{M}$ such that
\begin{equation}
\frac{\left| \left\langle\beta _{jk}\beta _{jk^{\prime
}}\right\rangle\right|}{\left\langle\beta _{jk}^2\right\rangle} \leq
\frac{C_{M}}{(1+B^{j}\arccos (|\xi _{jk}-\xi _{jk^{\prime
}}|))^{M}}\text{ .} \label{scacorrin}
\end{equation}
\end{itemize}

Property P1 is a straightforward consequence of the localization properties
for $b(x)$ in the harmonic domain. Property P2 is much more surprising, and
does not follow by any means from the localization in pixel space (\ref%
{expine}); indeed it is simple to provide examples of wavelet systems that
satisfy (\ref{expine}), and still do not enjoy (\ref%
{scacorrin}) (see \cite{lan2},\cite{Mayeli}). Both these properties hold for
any isotropic random field, without any assumption on its distribution (i.e.
Gaussianity).

Properties P1,P2 suggest that at high frequency, needlet coefficients can
be approximated as a sample of identically distributed and uncorrelated
(independent, in the Gaussian case) coefficients, and this property opens
the way to a huge toolbox of statistical procedures for CMB data analysis.
In practice, numerical approximations and the presence of masked regions
will entail that P1 and P2 will only hold approximately; nevertheless,
simulations have suggested that these properties do ensure a remarkable
performance of needlets when applied to actual data from CMB experiments,
see \cite{denest, gmcw, gm2, gm1} for further developments.

\section{Spin Needlets}

Throughout this paper, we shall assume that there exist a grid of \emph{%
cubature points} $\left\{ \xi _{jk}\right\} ,$ and a set of corresponding
weights $\left\{ \lambda _{jk}\right\} $ such that following discrete
approximations of spherical integrals hold:%
\begin{eqnarray}
\sum_{k}\lambda _{jk}\left\{ _{2}Y_{\ell m}(\xi _{jk})\right\} \overline{%
\left\{ _{2}Y_{\ell' m'}(\xi _{jk})\right\} } &\simeq &\int_{S^{2}}\left\{
_{2}Y_{\ell m}(\hat{\gamma})\right\} \overline{\left\{ _{2}Y_{\ell' m'}(\hat{%
\gamma})\right\} }d\hat{\gamma}  \label{cubpoi} \\
&=&\delta _{l}^{l^{\prime }}\delta _{m}^{m^{\prime }}\text{ .}  \notag
\end{eqnarray}%
Here $_{\pm 2}Y_{\ell m}$ are the spin spherical harmonics defined in the
appendix. For the standard scalar case, the existence of such points is
well-known and provided by many different constructions, see for instance
(\cite{npw1,npw2}) and (\cite{glesp}), see also \cite{bkmpAoS,bkmpBer}
for further discussion and \cite{gmcw,gm2,gm1} for extensions to the generalized
needlets case. For the spin spherical harmonics we consider here,
the validity of (\ref{cubpoi}) is
going to be investigated mathematically elsewhere. We stress, however, that
even if equation (\ref{cubpoi}) is not known, or the points $\left\{ \xi
_{jk}\right\} $ are not explicitly given, then one can use any sensible
collection of points and weights, and the results will still hold approximately; in other
words, our results below will continue to hold with minor numerical
approximations when implemented on any package with a pixelization scheme
such as HEALPix (this issue is discussed rigorously and in greater detail in %
\cite{spin-mat}).

Spin needlets are defined as (see (\cite{spin-mat}) for a complete
mathematical treatment and more rigorous results)%
\begin{equation}
\psi _{jk;2}(\hat{\gamma})=\sqrt{\lambda _{jk}}\sum_{\ell }b(\frac{\ell }{%
B^{j}})\sum_{m=-\ell }^{\ell }\left\{ _{2}Y_{\ell m}(\hat{\gamma})\right\}
\left\{ \overline{_{2}Y_{\ell m}(\xi _{jk})}\right\} \text{ .}
\label{spinwav2}
\end{equation}%
A comparison between (\ref{needlets_expansion}) and (\ref{spinwav2})
highlights immediately that spin needlets make up a natural extension of the
ideas underlying the approach in the scalar case to the framework of a spin
field. This deep link between the two constructions should not hide,
however, some profound differences between $\psi _{jk}(\hat{\gamma})$ and $%
\psi _{jk;2}(\hat{\gamma})$ as mathematical objects. Indeed, as recalled in
the previous Section $\psi _{jk}(\hat{\gamma})$ is a standard scalar
function which induces a linear map (\ref{needproj}) leading from
$T(\hat{\gamma})\rightarrow \beta _{jk},$ i.e. from a scalar quantity
to a scalar quantity. On the contrary, $\psi _{jk;2}$ induces a linear map leading from spin 2
quantities to spin 2 wavelet coefficients. The quantities in (\ref%
{spinwav2}) depend on the choice of the coordinate system for $\hat{\gamma}$
and $\xi _{jk};$ these two coordinate systems may be chosen independently.
If the coordinate system for $\hat{\gamma}$ is rotated, $\psi _{jk;2}(\hat{%
\gamma})$ transforms like a spin 2 vector at $\hat{\gamma},$ while if the
coordinate system at $\xi _{jk}$ is rotated, $\psi _{jk;2}(\hat{\gamma})$
transforms like a spin $-2$ vector at $\xi _{jk}.$ $\psi _{jk;2}(\hat{\gamma}%
)$ has a precise mathematical status (see (\cite{spin-mat})) as a linear map
from spin 2 vectors at $\xi _{jk}$ to spin two vectors at $\hat{\gamma}.$
Indeed, if $v$ is a spin 2 vector at $\xi _{jk}$, $\psi _{jk;2}(\hat{\gamma}%
)v$ makes sense as a spin 2 vector at $\hat{\gamma},$ since the product of a
spin $-2$ vector and a spin $2$ vector at a point is a well-defined complex
number, independent of choice of coordinates.
(Thus it would be more proper to write
$\left\{ _{2}Y_{\ell m}(\hat{\gamma})\right\} \otimes
\left\{ \overline{_{2}Y_{\ell m}(\xi _{jk})}\right\}$
than
$\left\{ _{2}Y_{\ell m}(\hat{\gamma})\right\}
\left\{ \overline{_{2}Y_{\ell m}(\xi _{jk})}\right\}$
in (\ref{spinwav2}).)

As we shall show below the fact that $\psi _{jk;2}(\hat{\gamma})$ is
not a scalar function does not prevent useful applications for the
reconstruction and testing on physically meaningful scalar
quantities such as the angular power spectra
$C_{l}^{EE},C_{l}^{BB}.$  We note
that in the mathematical results of \cite{spin-mat}, the factor $%
b(l/B^{j})$ is replaced by $b(\sqrt{(l-2)(l+3)}/B^{j});$ this reflects
the fact that the sequence $\left\{
(l-2)(l+3)\right\} _{l=3,4,...}$ represents the eigenvalues of the Laplacian
operator associated to spin spherical harmonics. However, since our
main interest here is high-frequency asymptotics, and since $l \sim \sqrt{(l-2)(l+3)}$
for large $l$, we use the simpler $b(l/B^{j})$ here
to highlight the similarity with the usual
presentation of scalar needlets. (Antithetically, one may modify the definition
of the latter by replacing $b(l/B^{j})$ with $b(\sqrt{l(l+1)}/B^{j});$ note
indeed that $\left\{ l(l+1)\right\} _{l=1,2}$ provides the sequence of
eigenvalues for the usual Laplacian operator on the sphere. We refer to \cite%
{gmcw,gm2,gm1} for more discussion on this point.)

\begin{figure}
\begin{center}
\leavevmode
\epsfig {file=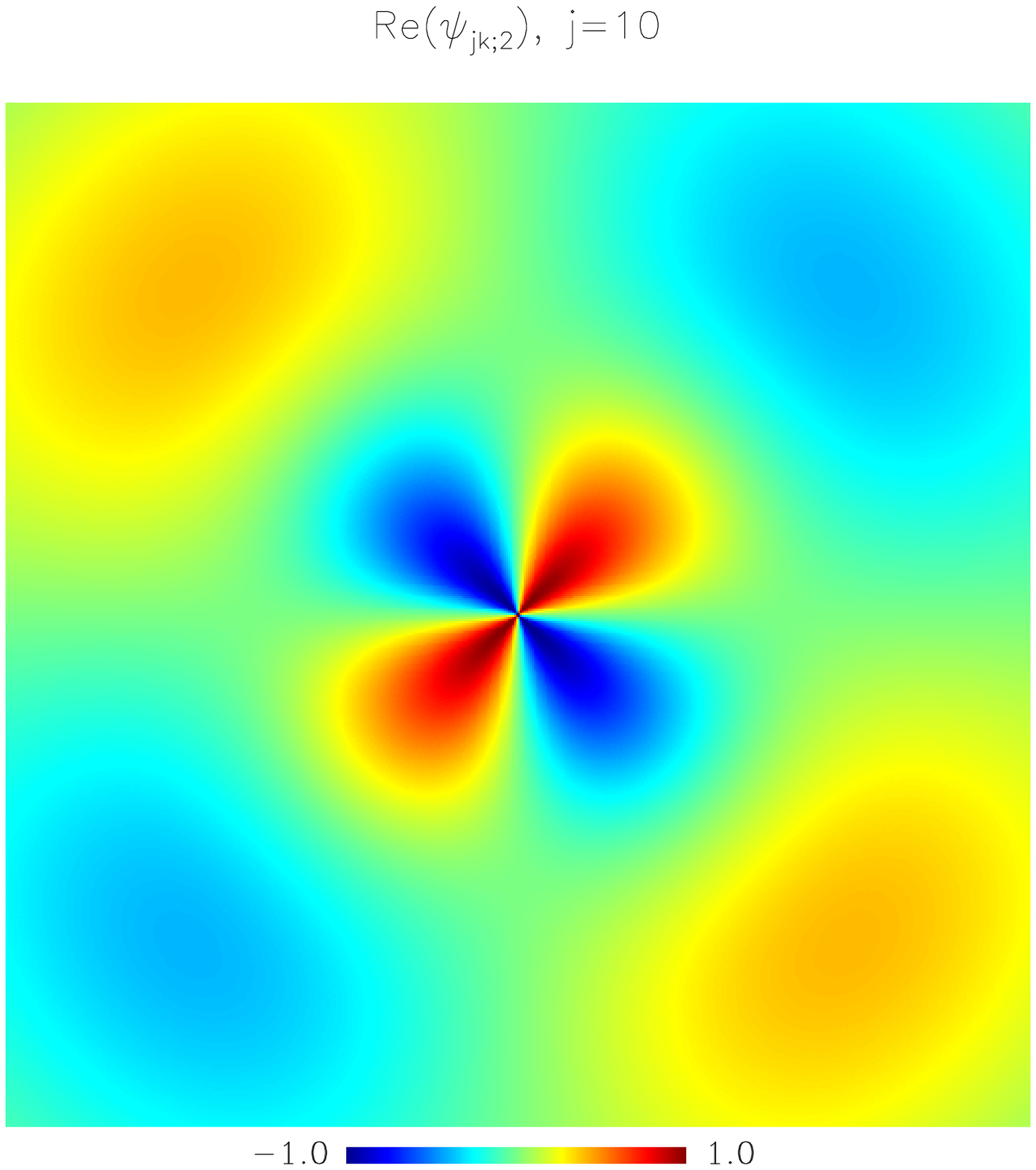,width=6cm,height=6cm}
\epsfig {file=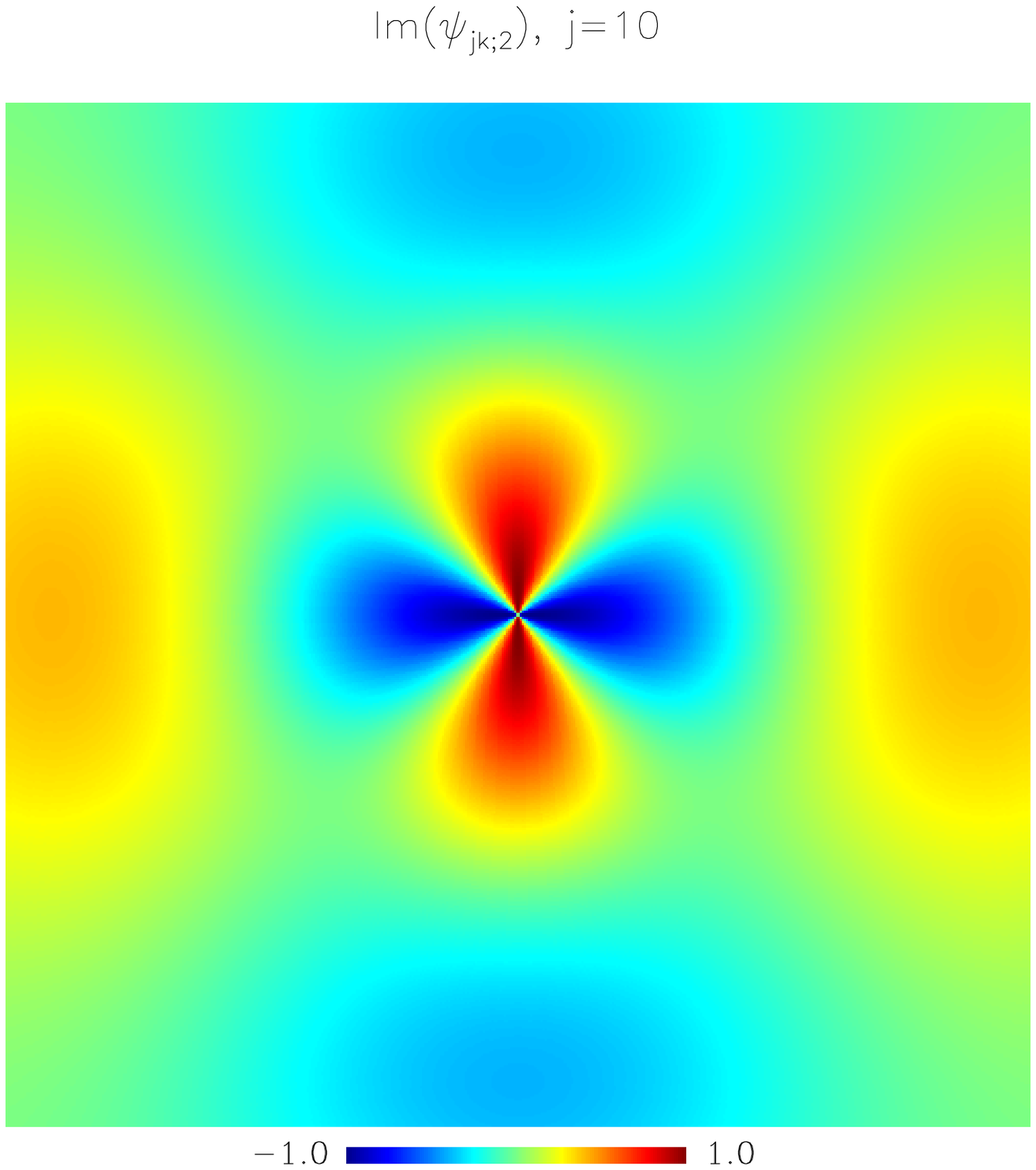,width=6cm,height=6cm}
\epsfig {file=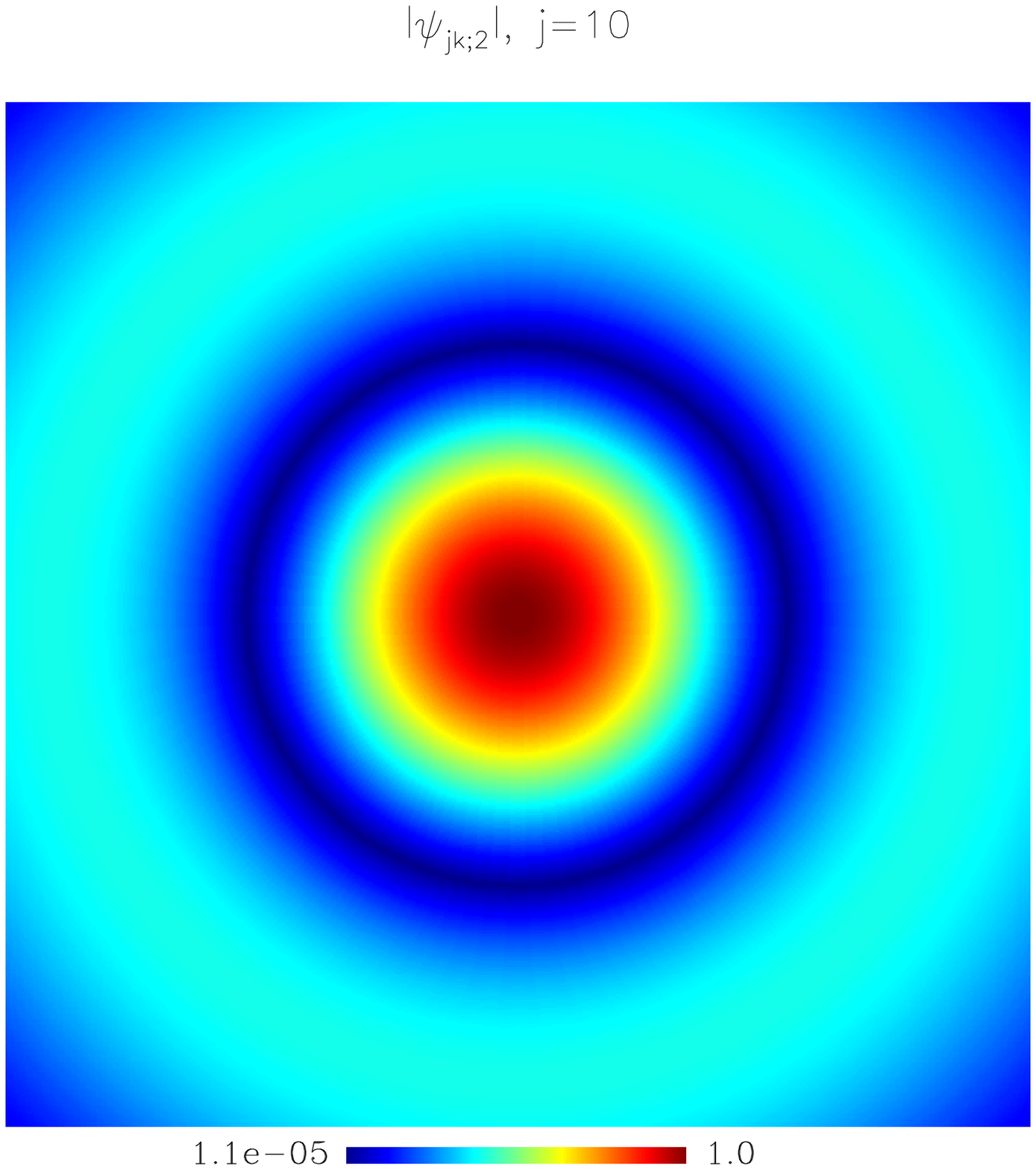,width=6cm,height=6cm}
\caption{Projections of $\psi _{jk;2}(\hat{\gamma})$ for $B=1.2$ and $j=10$. Left plot: real part, right plot: imaginary part, lower plot: modulus.}
\label{fig:gnom}
\end{center}
\end{figure}

\begin{figure}
\begin{center}
\leavevmode
\epsfig {file=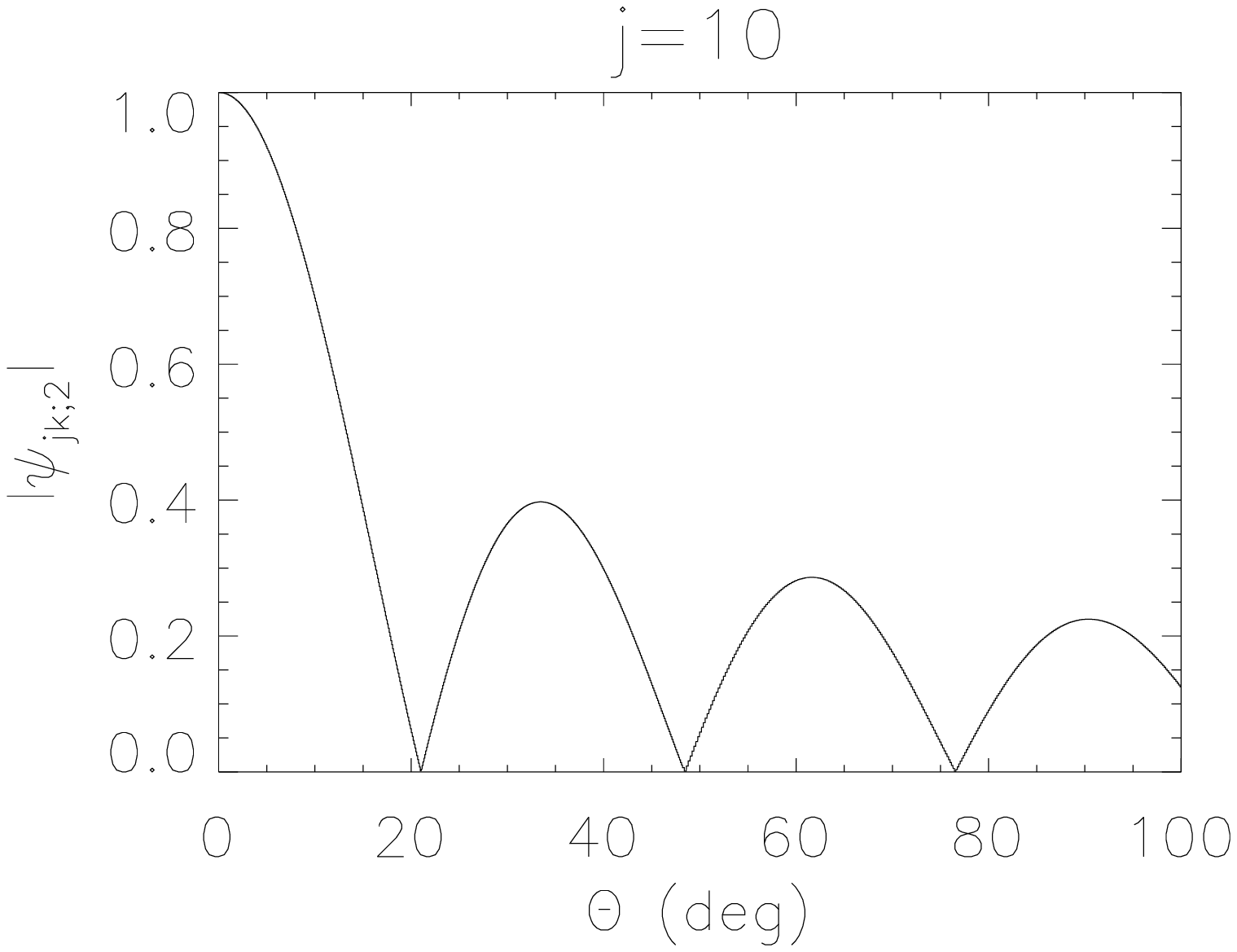,width=6cm,height=6cm}
\epsfig {file=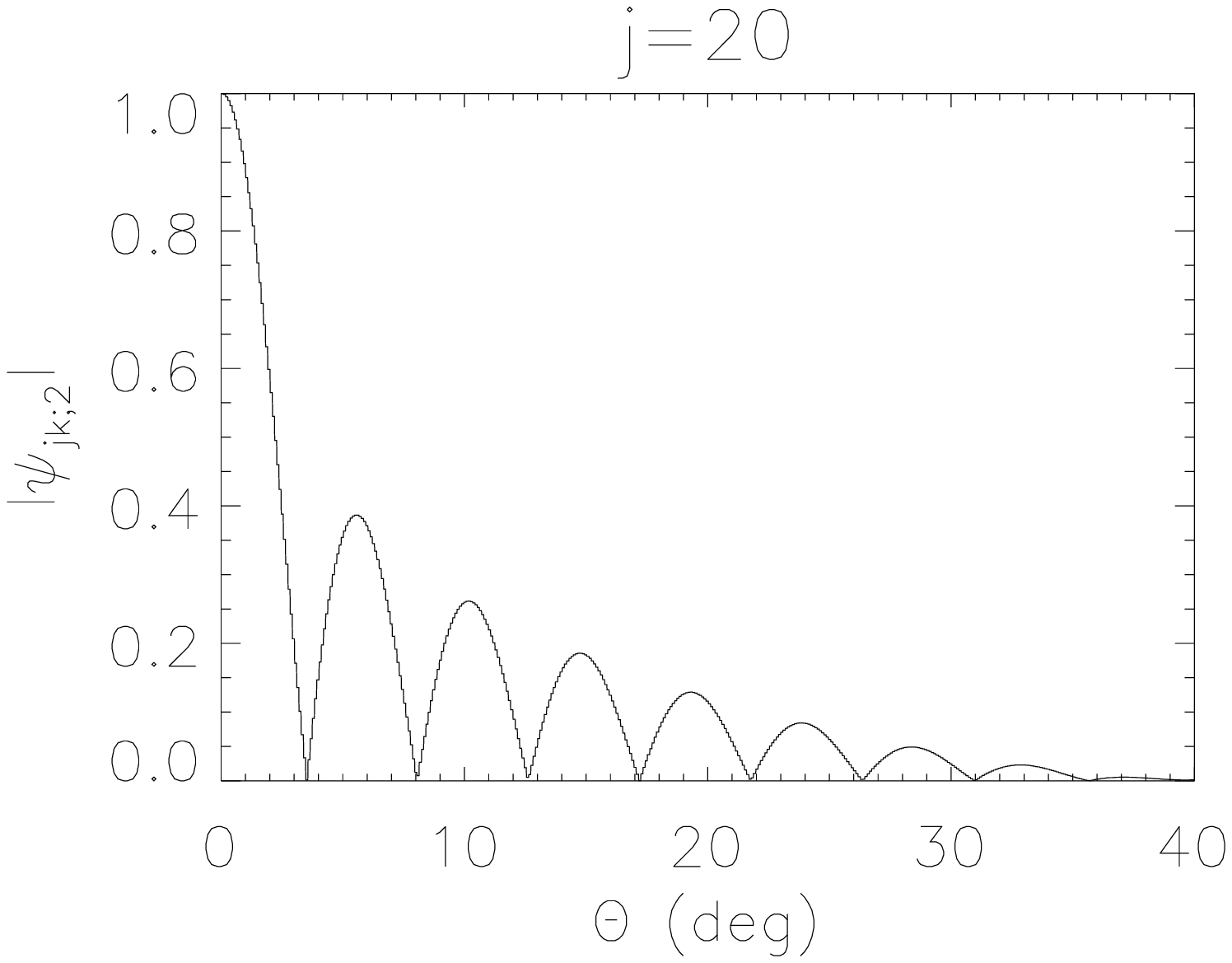,width=6cm,height=6cm}
\epsfig {file=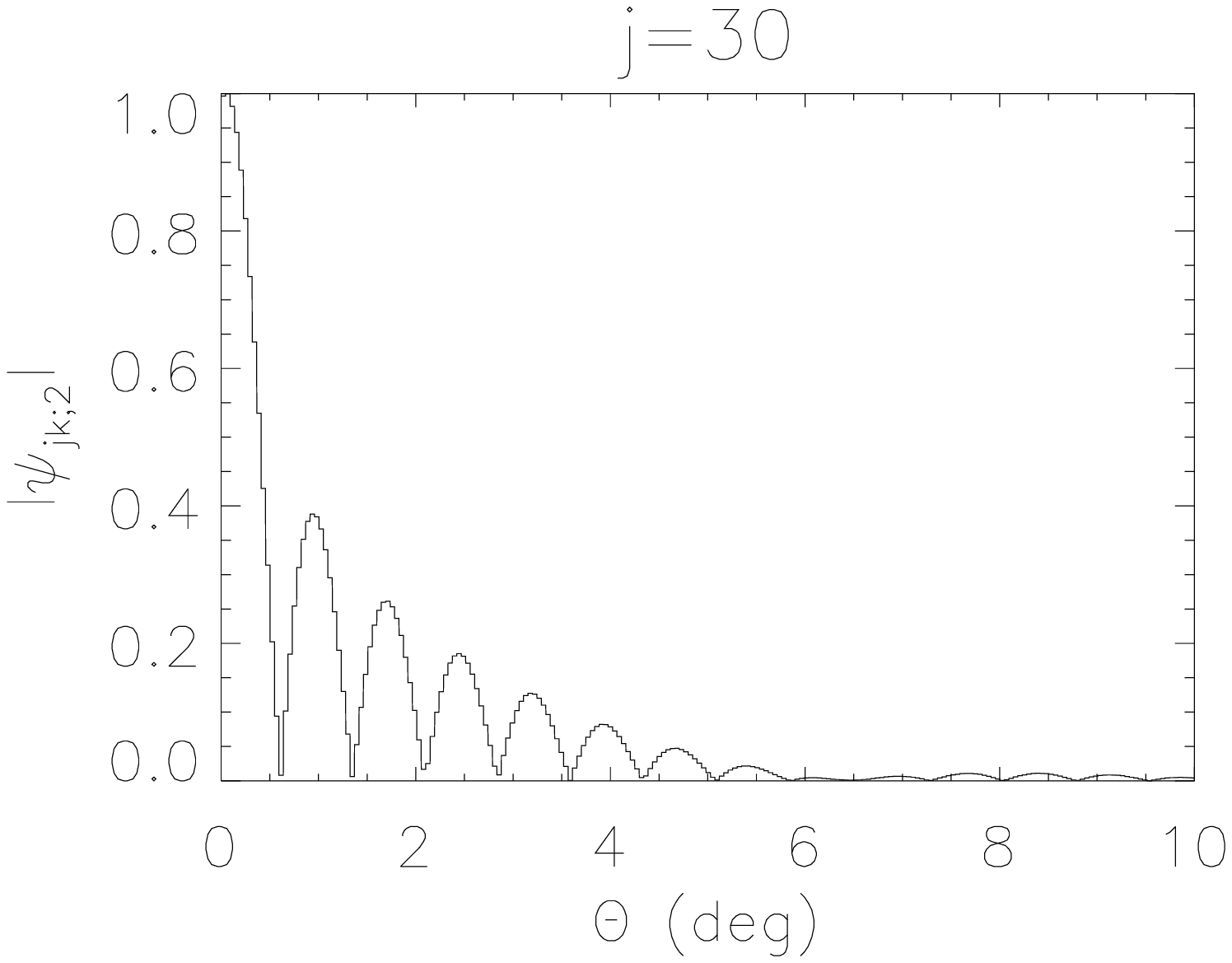,width=6cm,height=6cm}
\caption{The modulus of $\psi _{jk;2}(\hat{\gamma})$ for $B=1.2$ and $j=10$ (left plot), $j=20$ (right plot) and $j=30$ (lower plot). The angle $\theta$ is the distance from the center point.}
\label{fig:psi}
\end{center}
\end{figure}

In figure \ref{fig:gnom} we show the projection of $\psi _{jk;2}(\hat{\gamma})$ on the plane for $B=1.2$ and $j=10$. We show the real and imaginary component as well as the modulus. In figure \ref{fig:psi} we show how $\psi _{jk;2}(\hat{\gamma})$ falls off with the distance from the center for $B=1.2$ and $j=10,20,30$. As expected, we clearly see how $\psi _{jk;2}(\hat{\gamma})$ falls of faster for higher values of $j$ and thus picking up smaller scales.

We shall now investigate the localization properties of (\ref{spinwav2}).
Localization properties in multipole space are a straightforward consequence
of the properties of the weight function $b(x);$ localization properties in
pixel space are much less straightforward and require rather sophisticated
mathematical arguments. The following inequality is established in \cite%
{spin-mat} and generalizes (\ref{expine}):\\

{\bf Proposition 1}{\it \ \ \cite{spin-mat}\ \  For any $M=1,2,...$ there exists a positive constant $c_{M}$
such that for any point $\hat{\gamma}\in S^{2}$ we have%
\begin{equation}
|\psi _{jk;2}(\hat{\gamma})|\leq \frac{c_{M}B^{j}}{(1+B^{j}\arccos (|\hat{%
\gamma}-\xi _{jk}|))^{M}}\text{ .}  \label{spiine}
\end{equation}}

Note that although $\left\{ \psi _{jk;2}(\hat{\gamma})\right\} $ are \emph{%
not} scalar valued, $|\psi _{jk;2}(\hat{\gamma})|$ is clearly a well-defined
scalar function, so that the inequalities (\ref{spiine}) are consistent.

\begin{figure}
\begin{center}
\leavevmode
\epsfig {file=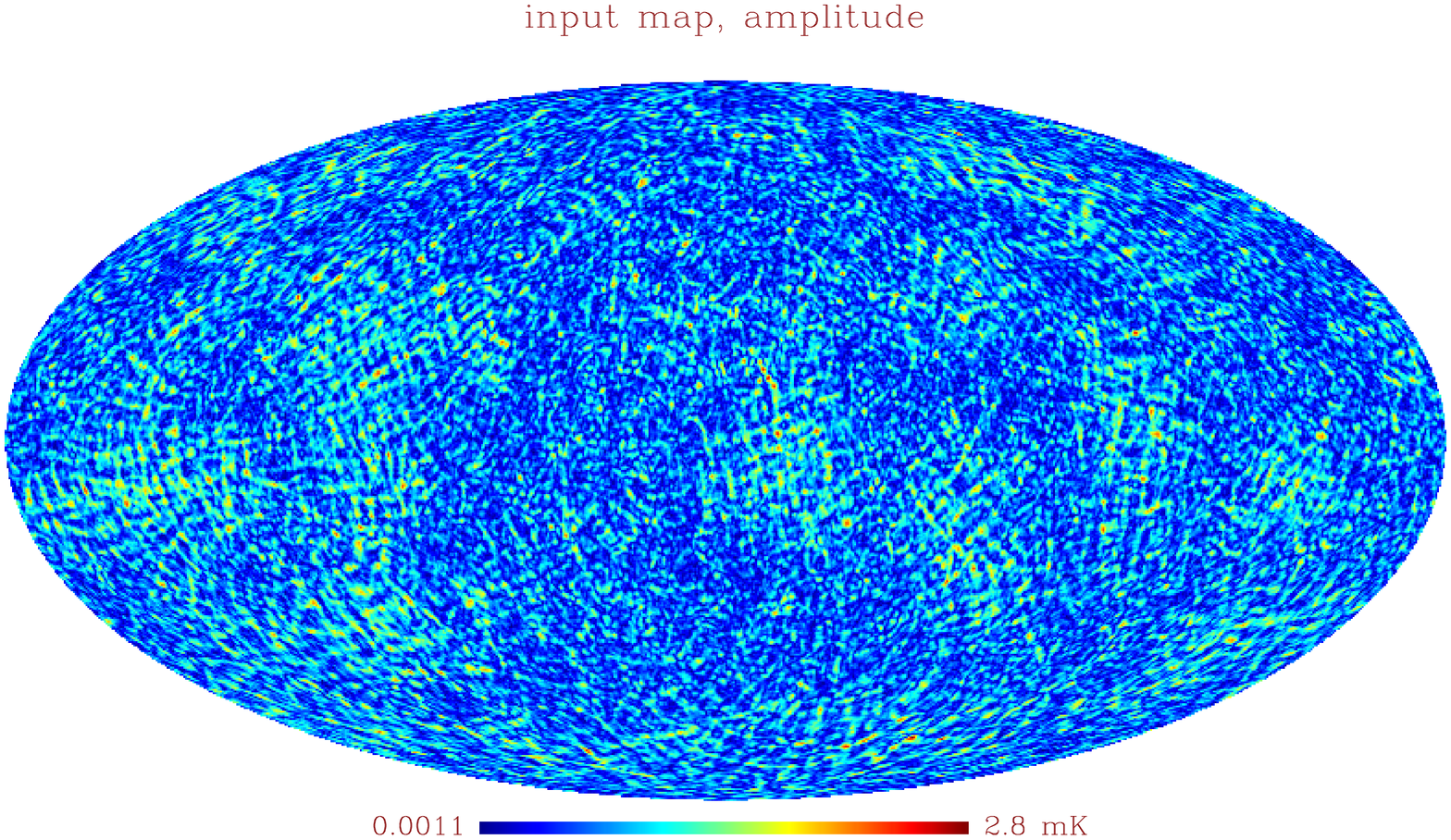,width=5cm,height=6cm,angle=90}
\epsfig {file=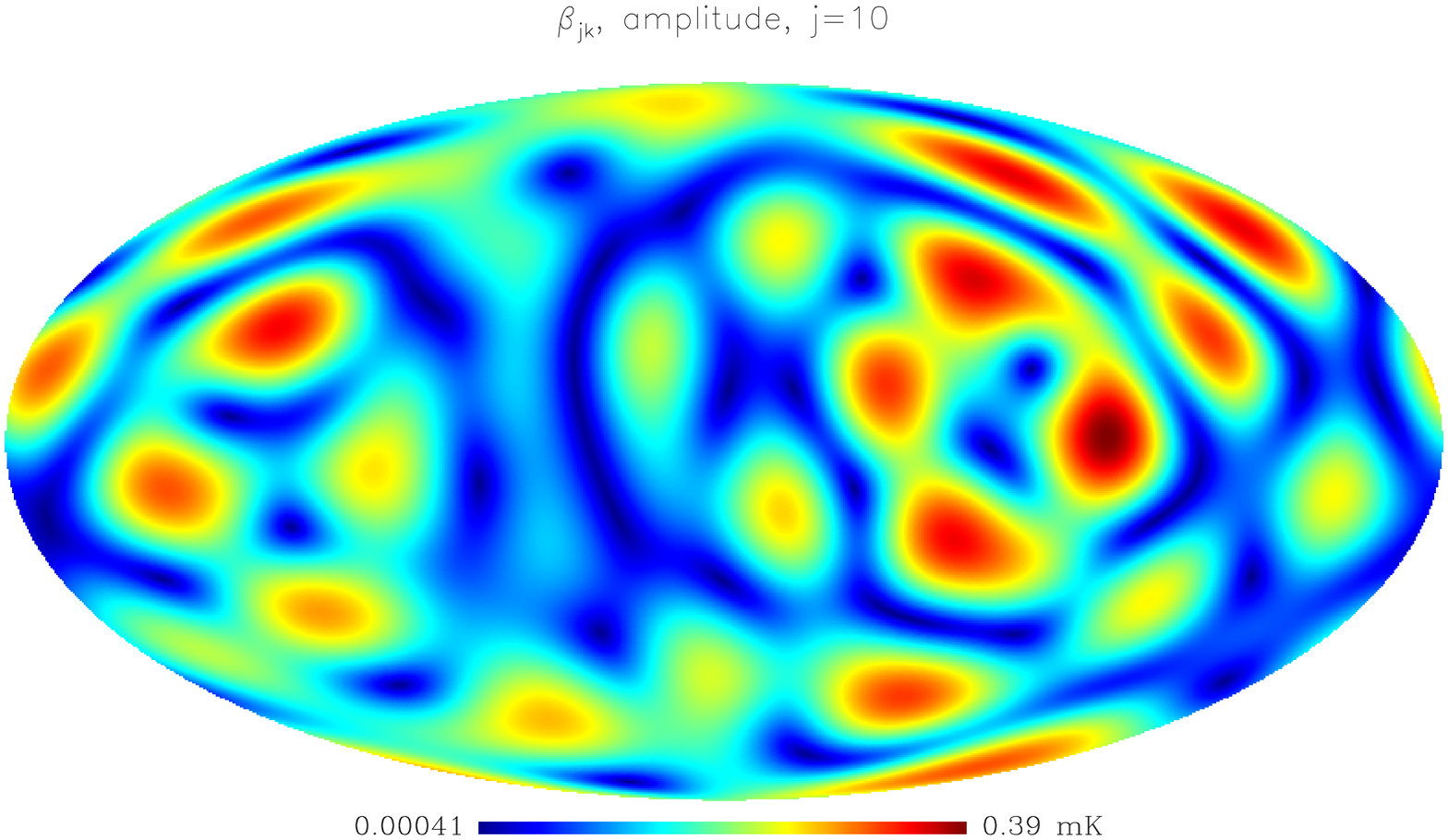,width=5cm,height=6cm,angle=90}
\epsfig {file=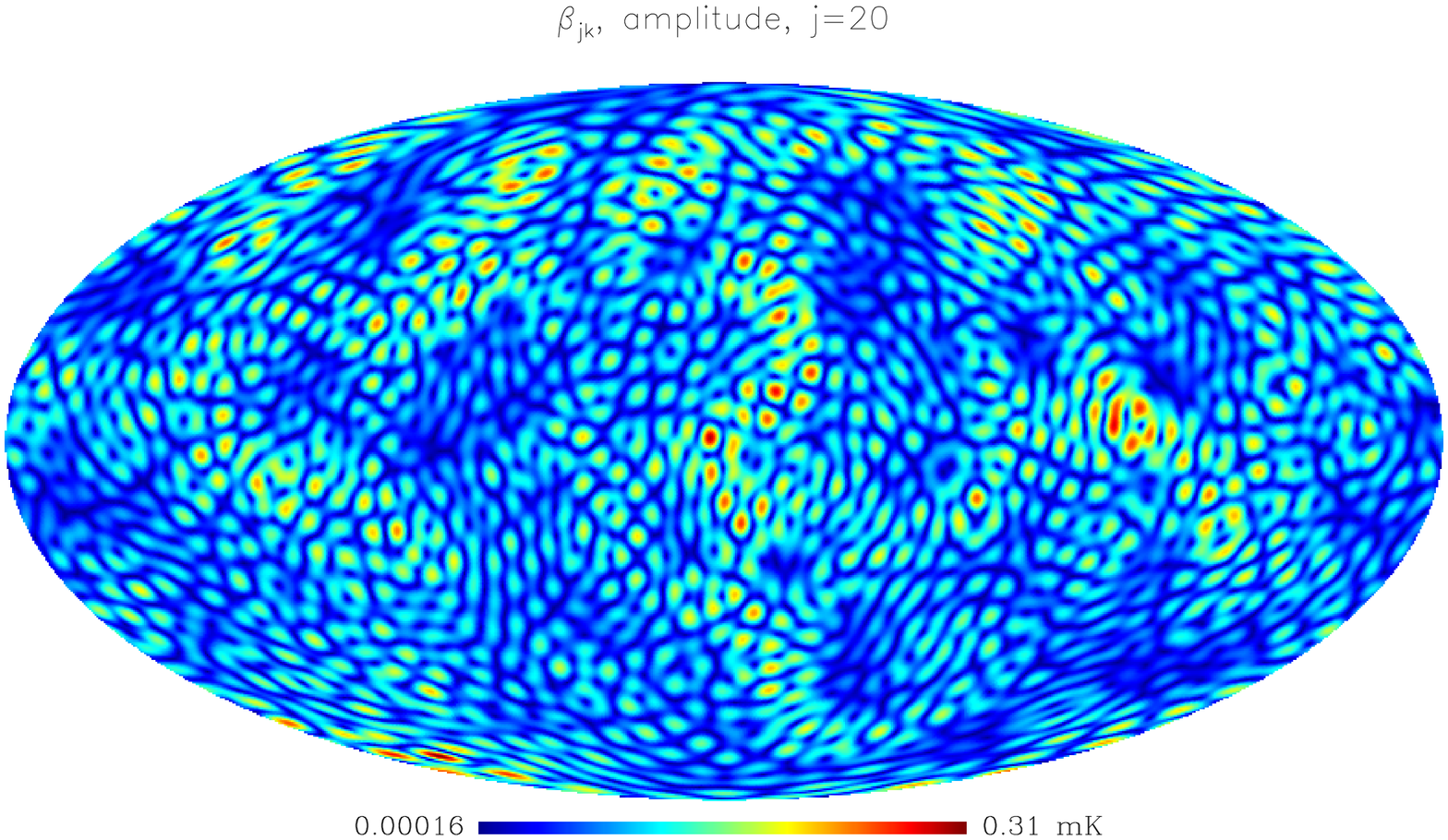,width=5cm,height=6cm,angle=90}
\epsfig {file=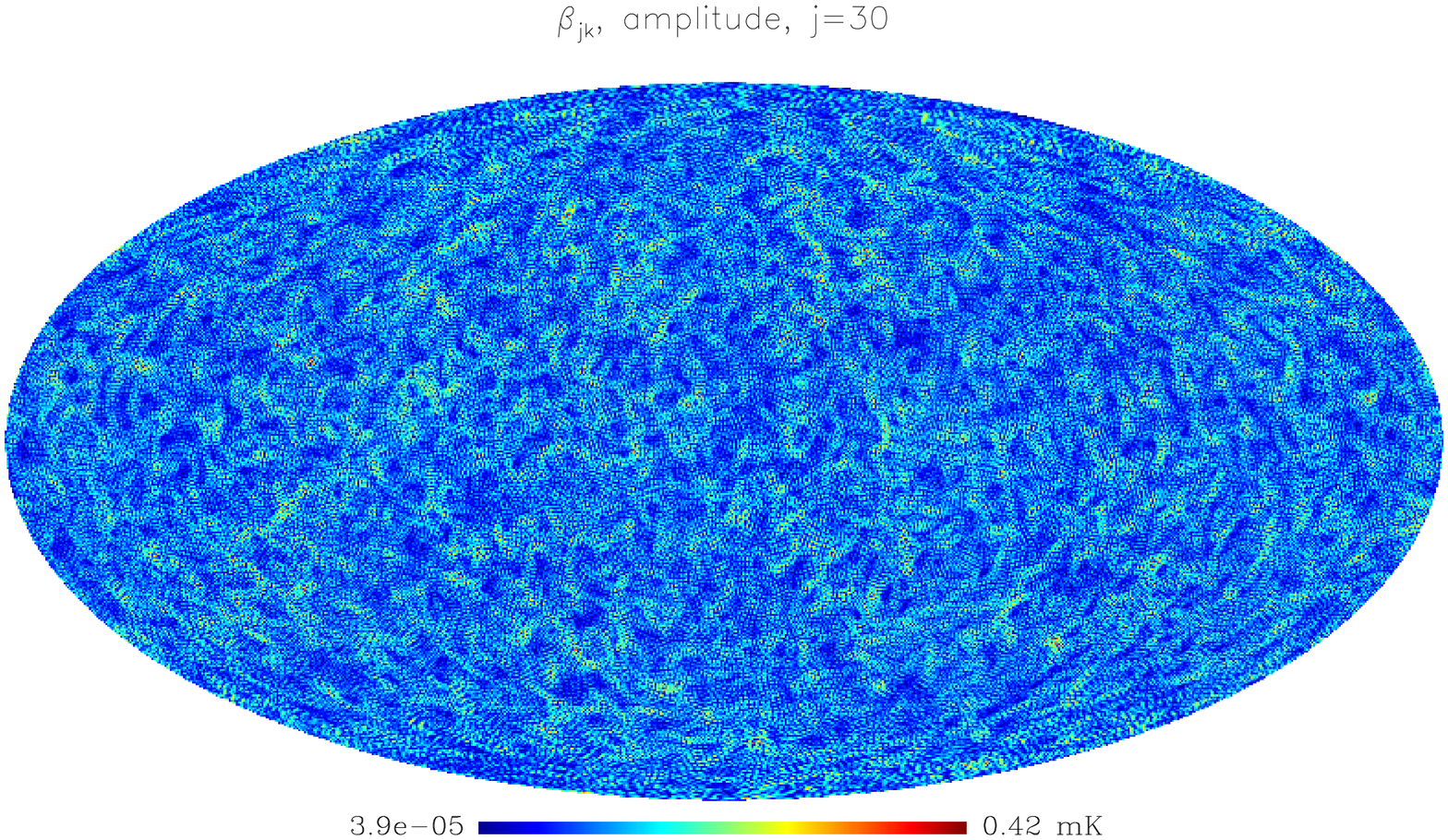,width=5cm,height=6cm,angle=90}
\caption{The input CMB map as well as the needlets coefficients for B=1.2 and j=10, 20 and 30. The plots show the polarization amplitude.}
\label{fig:beta_amp}
\end{center}
\end{figure}

\begin{figure}
\begin{center}
\leavevmode
\epsfig {file=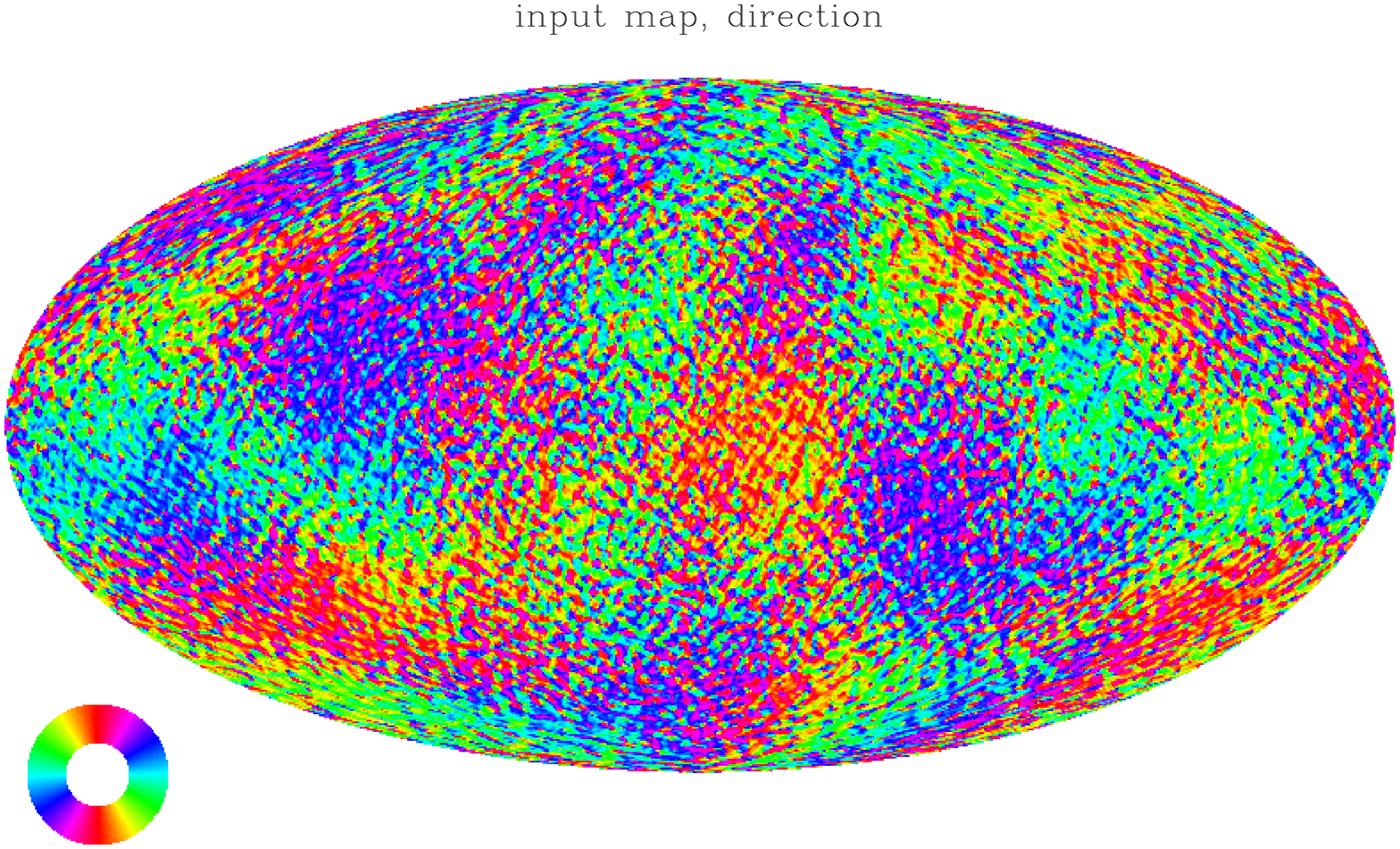,width=5cm,height=6cm,angle=90}
\epsfig {file=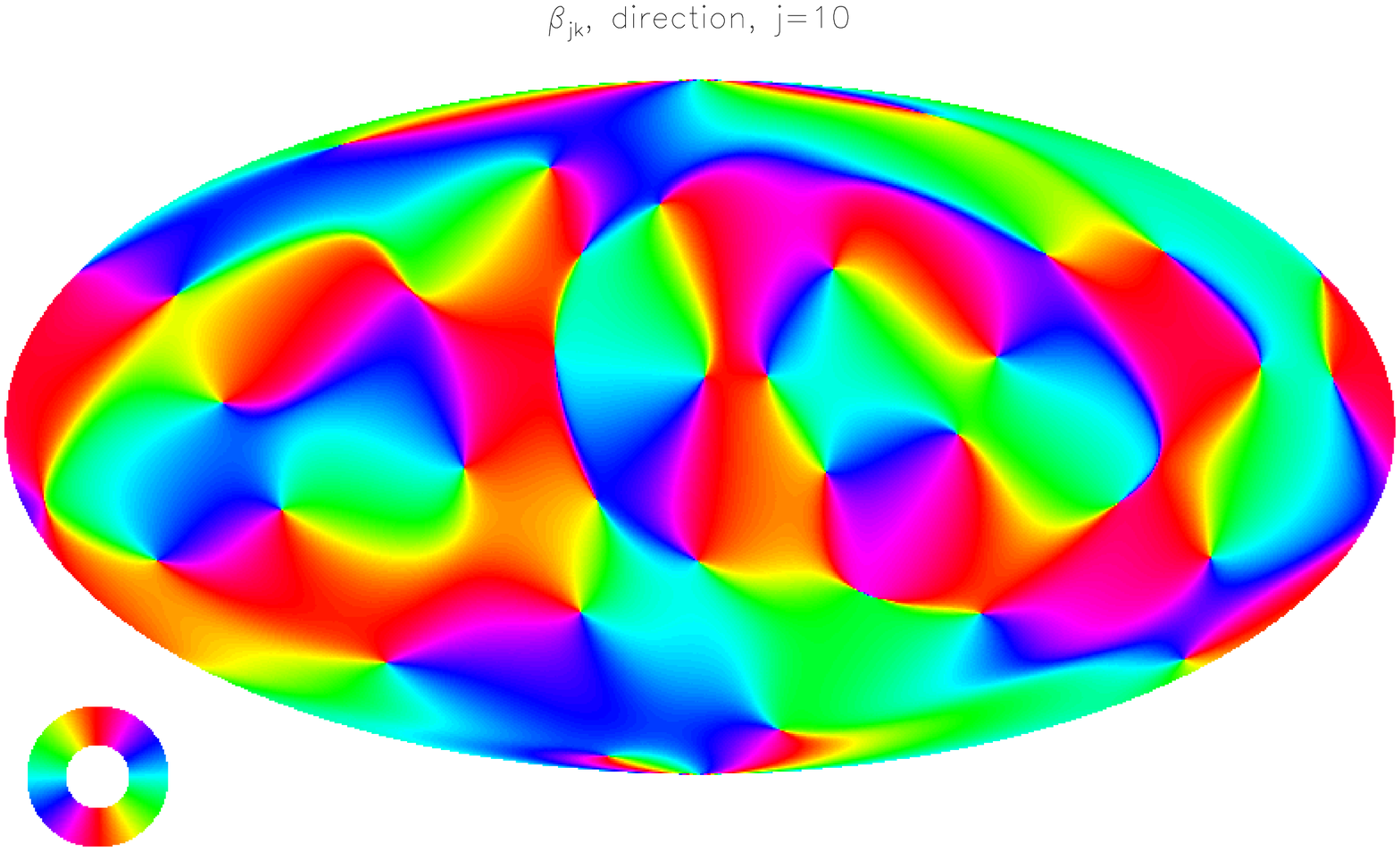,width=5cm,height=6cm,angle=90}
\epsfig {file=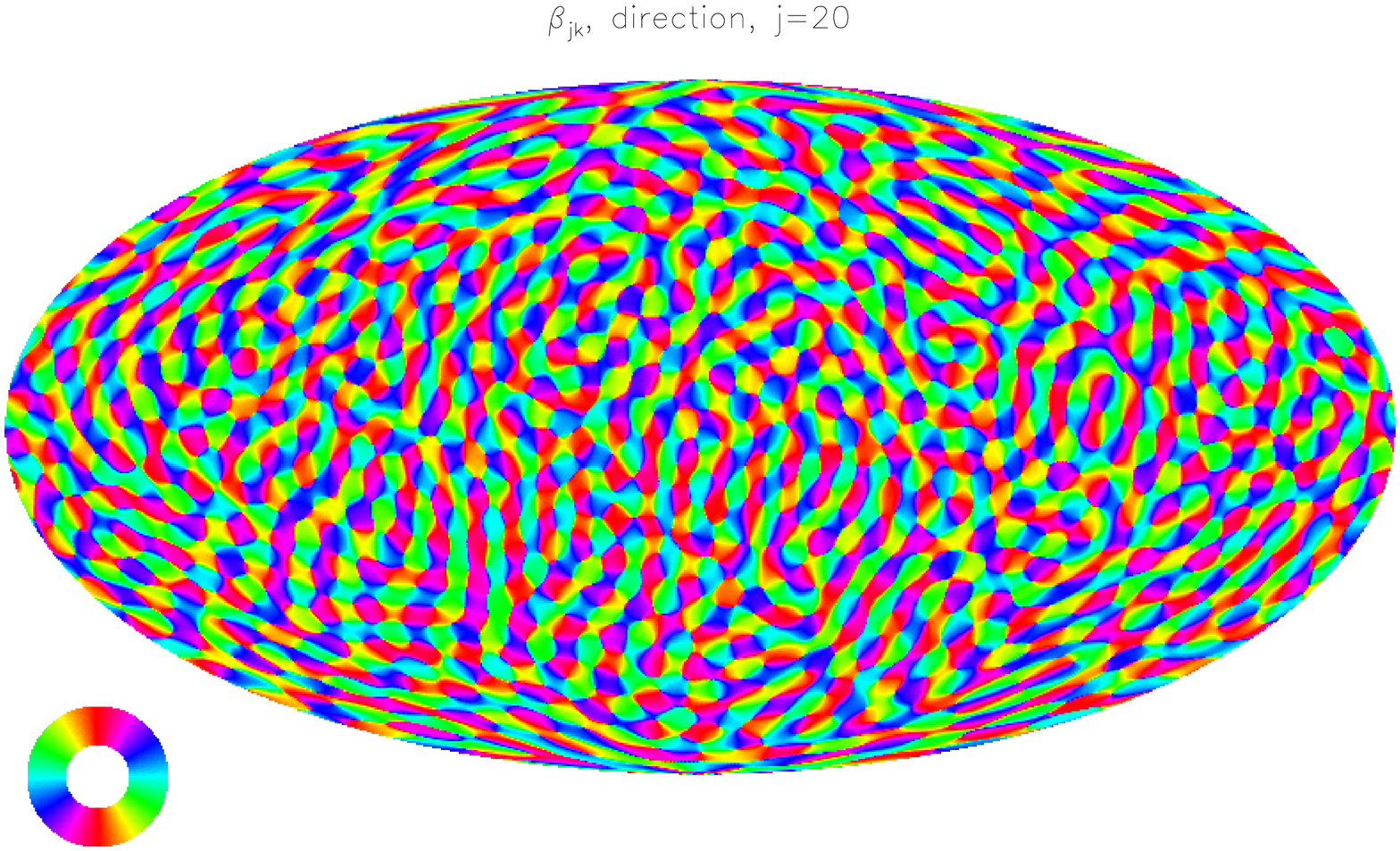,width=5cm,height=6cm,angle=90}
\epsfig {file=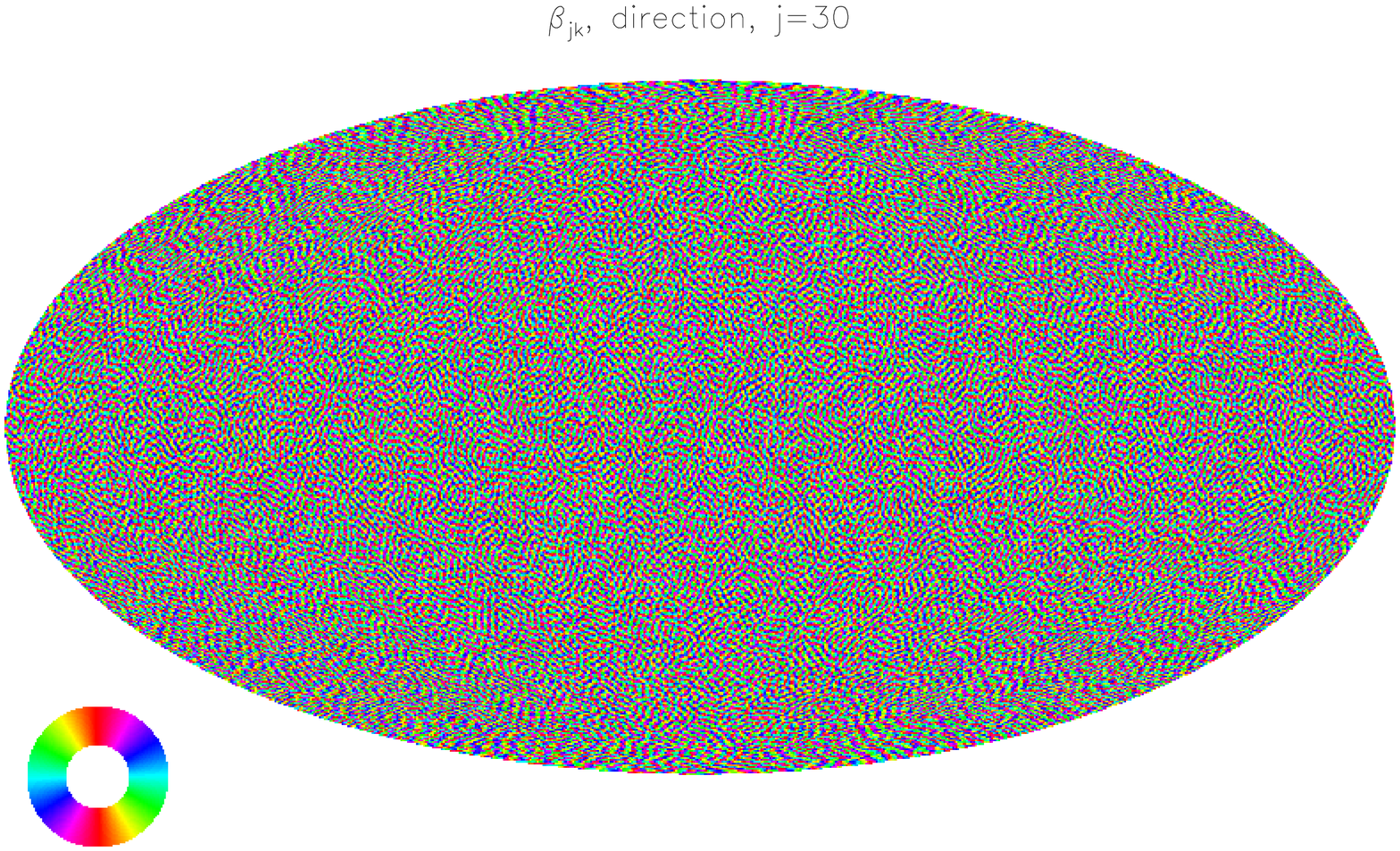,width=5cm,height=6cm,angle=90}
\caption{The input CMB map as well as the needlets coefficients for B=1.2 and j=10, 20 and 30. The plots show the polarization direction.}
\label{fig:beta_dir}
\end{center}
\end{figure}

Now consider the spin 2 fields
\begin{equation*}
Q(\hat{\gamma})+iU(\hat{\gamma})=\sum_{lm}a_{lm;2}\left\{ _{2}Y_{lm}(\hat{%
\gamma})\right\}
\end{equation*}%
where we have introduced the complex-valued random coefficients%
\begin{equation}
a_{lm;2}=\int_{S^{2}}\left\{ Q(\hat{\gamma})+iU(\hat{\gamma})\right\}
\overline{\left\{ _{2}Y_{lm}(\hat{\gamma})\right\} }d\hat{\gamma}%
=-(a_{lm;E}+ia_{lm;B})\text{ ,}  \label{four-spin}
\end{equation}%
where $a_{lm;E},a_{lm;B}$ denote, respectively, the spherical harmonics
coefficients of the $E$,$B$ components of the polarization random field. The
spin needlet coefficients are defined as
\begin{eqnarray*}
\beta _{jk;2} &:&=\int_{S^{2}}\left\{ Q(\hat{\gamma})+iU(\hat{\gamma}%
)\right\} \overline{\psi _{jk;2}(\hat{\gamma})}d\hat{\gamma} \\
&=&\sqrt{\lambda _{jk}}\sum_{lm}b(\frac{l}{B^{j}})a_{lm;2}\left\{
_{2}Y_{lm}(\xi _{jk})\right\} \text{ ,}
\end{eqnarray*}%
in obvious analogy to (\ref{needcof}) for the scalar case, only replacing
spin spherical harmonics and the corresponding (scalar) random coefficients$.
$ Note that
\begin{equation*}
\overline{\psi _{jk;2}(\hat{\gamma})}=\sqrt{\lambda _{jk}}\sum_{\ell }b(%
\frac{\ell }{B^{j}})\sum_{m=-\ell }^{\ell }\left\{ \overline{_{2}Y_{\ell m}(%
\hat{\gamma})}\right\} \left\{ _{2}Y_{\ell m}(\xi _{jk})\right\}
\end{equation*}%
is a spin $2$ vector at $\xi _{jk}$ and a spin $-2$ vector at $\hat{\gamma};$
$\beta _{jk;2}$ is a spin $2$ vector at $\xi _{jk}$. In other words, $Q(\hat{%
\gamma})+iU(\hat{\gamma})=\sum_{lm}a_{lm;2}\left\{ _{2}Y_{lm}(\hat{\gamma}%
)\right\} $ is spin $2;$ when multiplied with $\overline{\psi _{jk;2}(\hat{%
\gamma})},$ the spin $2$ factor annihilates with the spin $-2$ factor in $%
\overline{_{2}Y_{\ell m}(\hat{\gamma})},$ and we are just left with a spin $2
$ factor in $\beta _{jk;2},$ given by $\left\{ _{2}Y_{\ell m}(\xi
_{jk})\right\} .$ In figure \ref{fig:beta_amp} and  \ref{fig:beta_dir} we show maps of the amplitude and direction of polarization of the input map as well as for needlet coefficients at $j=10,20,30$ for $B=1.2$.

\begin{figure}
\begin{center}
\leavevmode
\epsfig {file=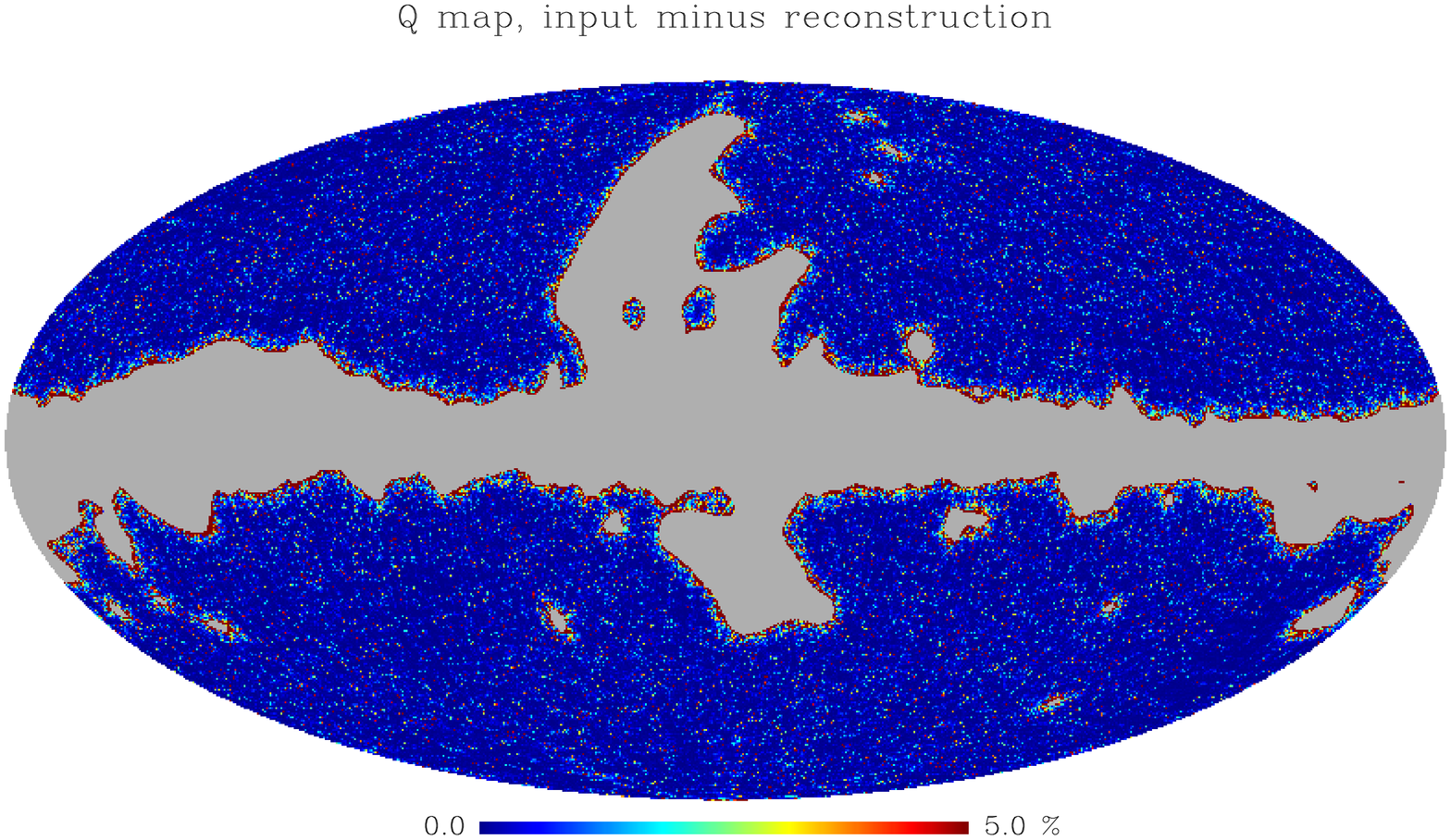,width=5cm,height=6cm,angle=90}
\epsfig {file=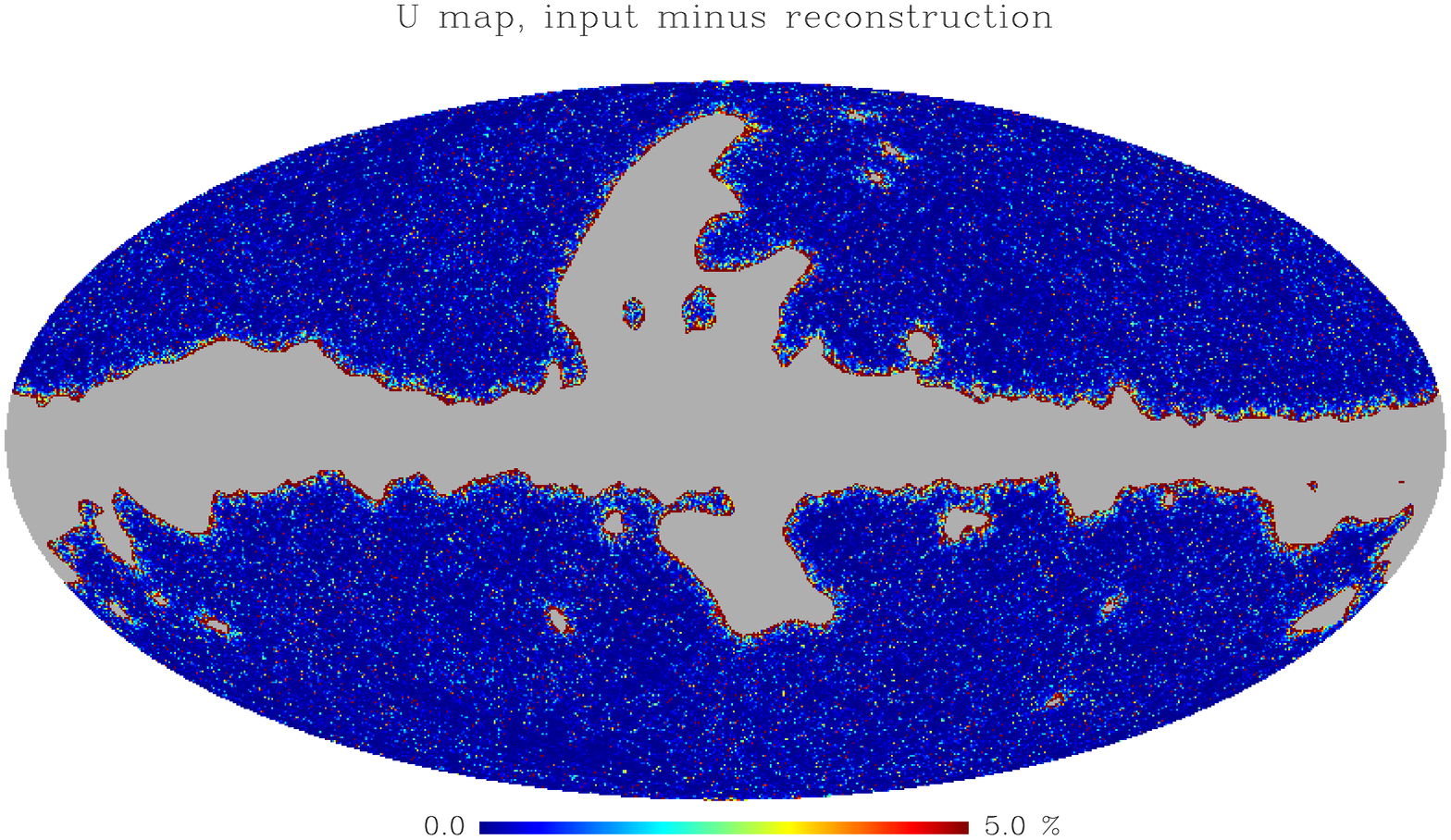,width=5cm,height=6cm,angle=90}
\caption{The relative difference (in $\%$ between an input CMB map (Q and U map) and the reconstructed map in the presence of a mask (in this case the P06 galactic cut used by the WMAP team for polarisation).}
\label{fig:reconstr}
\end{center}
\end{figure}

Assuming (\ref{cubpoi}), the following reconstruction formulae can be shown
to hold (see (\cite{spin-mat})

\begin{equation}
\left\{ Q(\hat{\gamma})+iU(\hat{\gamma})\right\} =\sum_{jk}\beta _{jk;2}\psi
_{jk;2}(\hat{\gamma})\text{ .}  \label{recfor1n}
\end{equation}%
Note that the right side of (\ref{recfor1n}) makes sense,
independent of coordinate system chosen for the $\xi _{jk}$, and
defines a spin $2$ vector at $\hat{\gamma}$. To see this, one need
only note again that the product of two quantities, one which
transforms like a spin $2$ vector at $\xi _{jk}$, and the other
which transforms like a spin $-2$ vector at $\xi _{jk}$, is a
well-defined complex number, independent of choice of coordinate
system. An example of the reconstruction is shown in figure
\ref{fig:reconstr}. In the figure we show the relative difference in
percent between the input and reconstructed Q and U maps. Outside
the masked regions, the difference is equal to the numerical noise
obtained by doing a simple spherical harmonic transform followed by
the inverse transform. Close to the masked regions, there is a small
area where the reconstruction error is larger. The size of this
region will depend on the choice of $B$. In this case $B=1.2$ which
is a rather low value resulting in a relatively large region with
larger error around the mask. Note that most of these points still
have reconstruction errors smaller than $5\%$. We will discuss the
accuracy of reconstruction in detail in a following paper.

As detailed in the previous Section, a capital property for the random
needlet coefficients in the scalar case is asymptotic uncorrelation at any
fixed angular distance (for a smooth angular power spectrum), as the
frequency parameter diverges ( $j\rightarrow \infty $)$.$ A natural question
is the extent to which this property may continue to hold in the spin case.
The answer to this question is provided in \cite{spin-mat}, where it is
shown that, under mild regularity conditions on the angular power spectra,
for all $M>0$ there exist positive constants $C_{M}$ such that%
\begin{equation}
\frac{\left\vert \left\langle \beta _{jk;2}\beta _{jk^{\prime
};2}\right\rangle \right\vert}{\left\langle \left\vert \beta _{jk;2}
\right\vert ^2\right\rangle} \leq \frac{C_{M}}{(1+B^{j}\arccos (|\xi
_{jk}-\xi _{jk^{\prime }}|))^{M}}\text{ .}  \label{spicorr}
\end{equation}%

Again, $\left\vert \left\langle \beta _{jk;2}\beta _{jk^{\prime
};2}\right\rangle \right\vert $ is a well defined scalar functions,
despite the fact that $\left\langle \beta _{jk;2}\beta _{jk^{\prime
};2}\right\rangle $ are spin quantities that depend on the choice of
coordinates for tangent planes. (\ref{spicorr}) suggests that the
spin spherical needlet coefficients can be consistently used for
angular power spectrum estimation and map reconstruction, as
detailed in the following
section; in the scalar case, this argument is rigorously derived in (\cite%
{bkmpAoS}), whereas for the spin situation we are concerned with here, we
refer again to \cite{spin-mat} for a more complete mathematical analysis.

Now assume that we have a masked region $M$ where no polarization data are
actually available; of course, this implies that we shall only be able to
recover the coefficients%
\begin{equation}
\widetilde{a}_{lm;2}=\int_{S^{2}/M}\left\{ Q(\hat{\gamma})+iU(\hat{\gamma}%
)\right\} \overline{\left\{ _{2}Y_{lm}(\hat{\gamma})\right\} }d\hat{\gamma}%
\text{ .}  \label{maskfc}
\end{equation}%
Due to the poor localization properties of spherical harmonics, the
coefficients $\left\{ \widetilde{a}_{lm;2}\right\} $ need not be close in
any meaningful sense to $\left\{ a_{lm;2}\right\} ;$ consequently any
statistical procedure based naively upon them (including estimation of $%
C_{l}^{EE},C_{l}^{BB}$ or the reconstruction of the scalar maps $E(\widehat{%
\gamma }),B(\widehat{\gamma })$) is likely to be seriously flawed. On the
other hand, consider%
\begin{equation*}
\widetilde{\beta }_{jk;2}=\sqrt{\lambda _{jk}}\sum_{lm}b(\frac{l}{B^{j}})%
\widetilde{a}_{lm;2}\left\{ _{2}Y_{lm}(\xi _{jk})\right\} \text{ ;}
\end{equation*}%
for locations $\left\{ \xi _{jk}\right\} $ that are ``sufficiently
away''\ from the masked region $M,$ i.e. $d(\xi _{jk},M)>\delta >0,$
it is easy to see that (see (\ref{spiine}))
\begin{eqnarray}
\left\langle \left| \widetilde{\beta }_{jk;2}-\beta _{jk;2}\right|
\right\rangle  &=&\left\langle \left| \int_{M}\left\{ Q(\hat{\gamma})+iU(%
\hat{\gamma})\right\} \overline{\psi _{jk;2}(\hat{\gamma})}d\hat{\gamma}%
\right| \right\rangle   \notag \\
&\leq &\frac{c_{M}B^{j}}{(1+B^{j}\arccos (\delta ))^{M}}\left\langle
\int_{M}\left| \left\{ Q(\hat{\gamma})+iU(\hat{\gamma})\right\} \right| d%
\hat{\gamma}\right\rangle   \notag \\
&=&o(B^{-j(M-1)})\text{ ,}  \label{betapp}
\end{eqnarray}%
i.e. the coefficients $\widetilde{\beta }_{jk;2},\beta _{jk;2}$ become
asymptotically equivalent at high frequencies $j\rightarrow \infty .$

It should be noted that throughout this paper we have decided to adopt the
spin spherical harmonics formalism considered by \cite{selzad}. A completely
analogous result could have been obtained by taking combinations of $%
b(l/B^{j})$ with \emph{grad }and \emph{curl harmonics, }i.e. starting from the
approach of \cite{kks96}. We do not focus on this possibility here for
brevity's sake.

\section{Some statistical applications}

The purpose of this Section is to provide some examples of the statistical
applications which can be entertained by means of spin needlet coefficients.
Our purpose here is \emph{not} to provide recipes which are ready-to-use for
CMB data analysis - this certainly requires much more computational and
analytical work to take into account the features of CMB maps: foregrounds,
anisotropic noise, multichannel observations and many others. Our purpose
here is different: it is well-known how wavelets in the scalar case have
proved to be a valuable instrument when dealing with a variety of
data analysis issues. The discussion below suggests that spin needlets can
be just as important.  Due to their specific nature as mathematical (spin)
objects, these applications however require extra care, even in the
idealistic circumstances we focus on here.

\subsection{Estimation of the sum of the angular power spectra $%
C_{l}^{EE}+C_{l}^{BB}$}

We recall that an estimator of the binned temperature power spectrum can be
constructed from scalar needlets, as follows. Take%
\begin{equation*}
\widehat{\Gamma }_{j}\overset{def}{=}\sum_{k}\beta _{jk}^{2}\text{ ;}
\end{equation*}%
in the idealistic case of no mask, it is readily seen that%
\begin{equation*}
\left\langle \widehat{\Gamma }_{j}\right\rangle =\sum_{k}\left\langle \beta
_{jk}^{2}\right\rangle =\sum_{B^{j-1}<l<B^{j+1}}b^{2}(\frac{l}{B^{j}}%
)(2l+1)C_{l}\overset{def}{=}\Gamma _{j}\text{ .}
\end{equation*}%
In the idealistic circumstances where the sky is fully-observed, the
previous sum includes $N_{j}$ terms, where we recall that $N_{j}$ is the
number of pixels at the resolution $j.$ The result continues to hold (with
some corrections to take into account the fraction of sky coverage) in the
presence of a sky cut, although clearly $\sum_{k}$ runs over a smaller
number of terms (in other words, a sky-fraction correction factor must be
introduced). This estimator was proposed in (\cite{bkmpAoS,pbm06}); in \cite%
{pbm06} applications are also shown to the cross-correlation between CMB\
and Large Scale Structure data, in \cite{fay08,fg08} this estimator is
extended to allow for the presence of observational noise, whereas in \cite%
{pietrobon08} the same approach is applied to search for features and
asymmetries in CMB maps. In (\cite{bkmpAoS}) it is also shown that, in the
Gaussian case%
\begin{equation}
\frac{\widehat{\Gamma }_{j}-\Gamma _{j}}{Var\left\{ \widehat{\Gamma }%
_{j}\right\} }\rightarrow _{d}N(0,1)\text{ ,}  \label{asyres}
\end{equation}%
where $\rightarrow _{d}$ denotes convergence in distribution and $N(0,1)$ a
standard Gaussian law. A result like (\ref{asyres}) provides also a clue for
testing goodness-of-fit and driving confidence intervals (at high
frequencies) for the angular power spectra. In \cite{bkmpBer} the behaviour
of statistics such as $\widehat{\Gamma }_{j}$ are considered for partial
regions of the sky, also as a device for testing asymmetries. From the
mathematical point of view, results like (\ref{asyres}) are entirely
justifiable on the basis of the uncorrelation properties we discussed
earlier in P1-P2 (see (\ref{scacorrin})). As the same sort of uncorrelation
property has been established in (\ref{spicorr}), it is natural to investigate
whether a similar procedure for the estimation of the binned angular power
spectra is feasible here.

The answer turns out to be positive, in the following sense. Consider the
(scalar valued) estimator%
\begin{equation*}
\widehat{\Gamma }_{j;2}=\sum_{k}\left| \beta _{jk;2}\right| ^{2}\text{ .}
\end{equation*}%
Starting from the idealistic case of no sky-cuts, we obtain easily%
\begin{eqnarray*}
\left\langle \widehat{\Gamma }_{j;2}\right\rangle  &=&\sum_{k}\left\langle
\left| \beta _{jk;2}\right| ^{2}\right\rangle \\
&=&\sum_{k}\left\langle \left| \sqrt{\lambda _{jk}}\sum_{lm}b(\frac{l}{B^{j}}%
)a_{lm;2}\left\{ _{2}Y_{lm}(\xi _{jk})\right\} \right| ^{2}\right\rangle  \\
&=&\sum_{k}\lambda _{jk}\sum_{l}b^{2}(\frac{l}{B^{j}})\Gamma
_{l}\sum_{m}\left| \left\{ _{2}Y_{lm}(\xi _{jk})\right\} \right| ^{2} \\
&=&\sum_{l}b^{2}(\frac{l}{B^{j}})\Gamma _{l}(2l+1)\overset{def}{=}\Gamma
_{j;2}\text{ ,}
\end{eqnarray*}%
because%
\begin{equation*}
\sum_{m}\left| \left\{ _{2}Y_{lm}(\xi _{jk})\right\} \right| ^{2}=\frac{2l+1%
}{4\pi }\text{ ,}
\end{equation*}%
and as before we took $\lambda _{jk}=4\pi /N_{j}.$ Here
\begin{equation*}
\Gamma _{l}=<\left| a_{lm;2}\right| ^{2}>=<\left| a_{lmE}+ia_{lmB}\right|
^{2}>=C_{l}^{EE}+C_{l}^{BB}.
\end{equation*}%
Likewise, it is possible to show that (\cite{spin-mat})%
\begin{equation}
\frac{\widehat{\Gamma }_{j;2}-\Gamma _{j;2}}{Var\left\{ \widehat{\Gamma }%
_{j;2}\right\} }\rightarrow _{d}N(0,1)\text{ , as }j\rightarrow \infty \text{
,}  \label{cltspin}
\end{equation}%
i.e. it is possible to prove that $\widehat{\Gamma }_{j;2}$ is consistent
and asymptotically Gaussian around its expected value $\Gamma _{j;2}.$ Note
that, as the electric and magnetic components of the polarization field are
uncorrelated, the variance in the denominator is simply the sum of the
variances of the two scalar components. Likewise, the asymptotic theory can
be developed as in \cite{bkmpAoS, bkmpBer}. As mentioned before, the
presence of masked regions of the sky requires the introduction of a sky
coverage fraction. The presence of anisotropic noise is more interesting and
can be dealt with along the lines of (\cite{fg08,fay08}), i.e. by introducing
weighted, rather than simple, averages of the squared needlet coefficients.

The previous procedure allows one to estimate (a binned form of) the angular
power spectra $C_{l}^{EE}+C_{l}^{BB}.$ Although this could be sufficient for
some purposes, it is clear that for CMB applications we are interested, in
general, in the estimation of $C_{l}^{EE}$ and $C_{l}^{BB}$ as separate
quantities; this is an issue to which we shall come back later in this
Section.

\subsection{Testing for non-Gaussianity}

As a further statistical application, it is possible to consider the
investigation of non-Gaussianity in the joint law of temperature and
polarization data. A standard idea to focus on non-Gaussianity is to
consider the skewness and kurtosis of wavelet coefficients; for brevity, we
shall concentrate on the latter statistic. To normalize our coefficients, we
estimate their variance by
\begin{equation*}
\widehat{\sigma }_{j}^{2}\overset{def}{=}\frac{1}{N_{j}}\sum_{k}\left| \beta
_{jk;2}\right| ^{2}=\frac{\widehat{\Gamma }_{j}}{N_{j}}\text{ ,}
\end{equation*}%
where we define also%
\begin{equation*}
\sigma _{j}^{2}\overset{def}{=}<\widehat{\sigma }_{j}^{2}>=\frac{1}{N_{j}}%
\sum_{k}<\left| \beta _{jk;2}\right| ^{2}>\text{ .}
\end{equation*}%
As a consequence of (\ref{cltspin}) and along the same lines as in \cite%
{bkmpAoS}, it can be shown that
\begin{equation*}
p\lim_{j\rightarrow \infty }\frac{\widehat{\sigma }_{j}^{2}}{\sigma _{j}^{2}}%
=p\lim_{j\rightarrow \infty }\frac{\widehat{\Gamma }_{j}}{\Gamma _{j}}=1%
\text{ , }
\end{equation*}%
$p\lim_{j\rightarrow \infty }$ denoting as before convergence in
probability; loosely speaking, this is to say that the mean of the squared
spin needlet coefficients is a consistent estimator of their variance. We
focus then on the normalized coefficients
\begin{equation*}
\widehat{\beta }_{jk;2}\overset{def}{=}\frac{\beta _{jk;2}}{\widehat{\sigma }%
_{j}}\text{ ;}
\end{equation*}%
to test for the joint Gaussianity of the temperature and polarization
fields, we may consider for instance their kurtosis, i.e.%
\begin{equation*}
\widehat{K}_{j;PP}=\sum_{k}\left\{ \left| \widehat{\beta }_{jk;2}\right|
^{4}-3\right\} \text{ },
\end{equation*}%
which converges to zero in probability as $j\rightarrow \infty .$
More interestingly, we might be interested in testing for the joint
Gaussianity of polarization and temperature maps. Likewise, it seems
possible to focus on the needlet bispectrum for joint temperature
and polarization data, as suggested for the scalar case by
\cite{lan}. These issues will be developed in a future paper.

\subsection{Reconstruction of the $E$ and $B$ maps}

A common strategy to reconstruct the (scalar) maps of the electric
and magnetic (or grad and curl) components of the polarization field
is well-known; starting from the maps $\left\{ Q\pm iU\right\} ,$
the scalar coefficients are evaluated by means of the Fourier
inversions (\ref{four-spin}). The maps are then recovered by the
standard spectral expansion. As simple as this procedure can seem,
it is prone to severe problems in the analysis of actual data, where
the presence of masked regions can lead to severe errors in the
evaluation of exact Fourier transforms. We shall show here how one
can use needlets to produce scalar maps by a novel technique.

We recall first how to obtain scalar maps from polarization data; indeed, as
in (\cite{wiaux06}) we focus on%
\begin{equation*}
\widetilde{E}(\hat{\gamma})\equiv -\frac{1}{2}\left[ \left( \overline{%
\partial }\right) ^{2}(Q+iU)(\hat{\gamma})+\left( \partial \right)
^{2}(Q-iU)(\hat{\gamma})\right]
\end{equation*}%
where we have used the spin raising and spin lowering operators $(\partial ,%
\overline{\partial })$ defined in the appendix. The crucial property to
recall is
\begin{eqnarray*}
\left( \overline{\partial }\right) ^{2}\left\{ _{2}Y_{\ell m}(\widehat{%
\gamma })\right\}  &=&\sqrt{\frac{(\ell +2)!}{(\ell -2)!}}Y_{\ell m}(%
\widehat{\gamma })\text{ ,} \\
\left( \partial \right) ^{2}\left\{\overline{_{2}Y_{\ell m}}(\widehat{\gamma }%
)\right\}  &=&\sqrt{\frac{(\ell +2)!}{(\ell -2)!}}\overline{Y_{\ell
m}}(\widehat{\gamma })\text{ .}
\end{eqnarray*}%
Hence we obtain, in view of (\ref{recfor1n})
\begin{eqnarray}
\widetilde{E}(\hat{\gamma}) &=&-\frac{1}{2}\left[ \left( \overline{\partial }%
\right) ^{2}\sum_{jk}\beta _{jk;2}\psi _{jk;2}(\hat{\gamma})+\left( \partial
\right) ^{2}\sum_{jk}\overline{\beta _{jk;2}}\overline{\psi _{jk;2}}(\hat{%
\gamma})\right]   \notag \\
&=&-\frac{1}{2}\sum_{jk}\beta _{jk;2}\sum_{\ell m}b(%
\frac{\ell }{B^{j}})\sqrt{\frac{(\ell +2)!}{(\ell -2)!}}\left\{ Y_{\ell m}(%
\hat{\gamma})\right\} \left\{ \overline{_{2}Y_{\ell m}}(\xi _{jk})\right\}
\notag \\
&&-\frac{1}{2}\sum_{jk}\overline{\beta _{jk;2}}%
\sum_{\ell m}b(\frac{\ell }{B^{j}})\sqrt{\frac{(\ell +2)!}{(\ell -2)!}}%
\left\{ \overline{Y_{\ell m}}(\hat{\gamma})\right\} \left\{ _{2}Y_{\ell
m}(\xi _{jk})\right\} \text{ .}  \label{emap}
\end{eqnarray}%
\begin{equation*}
=-\frac{1}{2}\sum_{jk}\left\{ \beta _{jk;2}\varphi _{jk}(%
\hat{\gamma})+\overline{\beta _{jk;2}}\overline{\varphi _{jk}}(\hat{\gamma}%
)\right\} \text{ ,}
\end{equation*}%
where
\begin{eqnarray}
&&\varphi _{jk}(\hat{\gamma})\overset{def}{=}\sqrt{\lambda _{jk}}\sum_{\ell m}b(\frac{\ell }{%
B^{j}})\sqrt{\frac{(\ell +2)!}{(\ell -2)!}}\left\{ Y_{\ell m}(\hat{\gamma}%
)\right\} \left\{ \overline{_{2}Y_{\ell m}}(\xi _{jk})\right\}
\label{fidif} \\
&=&\left( \overline{\partial }\right) ^{2}\psi _{jk;2}(\hat{\gamma})\text{ ,}%
\label{fidif1}
\end{eqnarray}%
which is scalar in $\hat{\gamma},$ spin $-2$ in $\xi _{jk}.$ Likewise%
\begin{eqnarray}
\widetilde{B}(\hat{\gamma}) &\equiv &\frac{i}{2}\left[ \left( \overline{%
\partial }\right) ^{2}(Q+iU)(\hat{\gamma})-\left( \partial \right)
^{2}(Q-iU)(\hat{\gamma})\right]   \notag \\
&=&\frac{i}{2}\left[ \left( \overline{\partial }\right) ^{2}\sum_{jk}\beta
_{jk;2}\psi _{jk;2}(\hat{\gamma})+\left( \partial \right) ^{2}\sum_{jk}%
\overline{\beta _{jk;2}}\overline{\psi _{jk;2}}(\hat{\gamma})\right]   \notag
\\
&=&\frac{i}{2}\sum_{jk}\beta _{jk;2}\sum_{\ell m}b(\frac{%
\ell }{B^{j}})\sqrt{\frac{(\ell +2)!}{(\ell -2)!}}\left\{ Y_{\ell m}(\hat{%
\gamma})\right\} \left\{ \overline{_{2}Y_{\ell m}}(\xi _{jk})\right\}
\notag \\
&&-\frac{i}{2}\sum_{jk}\:\overline{\beta _{jk;2}}%
\sum_{\ell m}b(\frac{\ell }{B^{j}})\sqrt{\frac{(\ell +2)!}{(\ell -2)!}}%
\left\{ \overline{Y_{\ell m}}(\hat{\gamma})\right\} \left\{ _{2}Y_{\ell
m}(\xi _{jk})\right\}   \label{bmap}
\end{eqnarray}%
\begin{equation*}
=\frac{i}{2}\sum_{jk}\left\{ \beta _{jk;2}\varphi _{jk}(%
\hat{\gamma})-\overline{\beta _{jk;2}}\overline{\varphi _{jk}}(\hat{\gamma}%
)\right\} \text{ .}
\end{equation*}%
In the presence of fully observed maps, it is immediate that%
\begin{eqnarray*}
\widetilde{E}(\hat{\gamma}) &=&-\frac{1}{2}\sum_{jk}\sqrt{\lambda _{jk}}%
\left\{ \sum_{lm}\sqrt{\lambda _{jk}}b(\frac{l}{B^{j}})a_{lm;2}\left(
_{2}Y_{lm}(\xi _{jk})\right) \right\} \sum_{\ell m^{\prime }}b(\frac{\ell }{%
B^{j}})\sqrt{\frac{(\ell +2)!}{(\ell -2)!}}\left\{ Y_{\ell m^{\prime }}(\hat{%
\gamma})\right\} \left\{ \overline{_{2}Y_{\ell m^{\prime }}}(\xi
_{jk})\right\}  \\
&&-\frac{1}{2}\sum_{jk}\sqrt{\lambda _{jk}}\left\{ \sum_{lm}\sqrt{\lambda
_{jk}}b(\frac{l}{B^{j}})\overline{a}_{lm;2}\left( \overline{_{2}Y_{lm}}(\xi
_{jk})\right) \right\} \sum_{\ell m^{\prime }}b(\frac{\ell }{B^{j}})\sqrt{%
\frac{(\ell +2)!}{(\ell -2)!}}\left\{ \overline{Y_{\ell m^{\prime }}}(\hat{%
\gamma})\right\} \left\{ _{2}Y_{\ell m^{\prime }}(\xi _{jk})\right\}
\end{eqnarray*}%
\begin{eqnarray}
&=&-\frac{1}{2}\sum_{lm}(a_{lm;2}+\overline{a}_{l-m;2})Y_{\ell m^{\prime}}(\hat{\gamma%
})\sqrt{\frac{(\ell +2)!}{(\ell -2)!}}\sum_{j}b(\frac{l}{B^{j}})b(\frac{\ell
}{B^{j}})\delta _{l}^{\ell }\delta _{m}^{m\prime }  \notag \\
&=&\sum_{lm}a_{lm;E}\sqrt{\frac{(\ell +2)!}{(\ell -2)!}}Y_{lm}(\hat{\gamma})%
\text{ ;}  \label{enorm1}
\end{eqnarray}%
likewise, it is immediate that
\begin{equation}
\widetilde{B}(\hat{\gamma})=\sum_{lm}a_{lm;B}\sqrt{\frac{(l+2)!}{(l-2)!}}%
Y_{lm}(\hat{\gamma})\text{ ,}  \label{bnorm1}
\end{equation}%
as expected. This is just a rephrasing of (\ref{recfor1n}) in terms of the
underlying (\emph{electric and magnetic}) scalar fields.

In the presence of masked maps, $\ E$ and $B$ are unfeasible; our idea is
then to use thresholding techniques, which have been shown to be
powerful when combined with the wavelet approach (see for instance \cite%
{DIKP}, or \cite{fg08} for applications in a CMB framework). In particular,
assume a portion $M$ of $S^{2}$ is masked; for each pixel $\left\{ \xi
_{jk}\right\} $ we can define the fraction
\begin{equation*}
s_{jk}=\frac{\int_{S^{2}/M}\left| \psi _{jk;2}(\hat{\gamma})\right| ^{2}d%
\hat{\gamma}}{\int_{S^{2}}\left| \psi _{jk;2}(\hat{\gamma})\right| ^{2}d\hat{%
\gamma}}\text{ .}
\end{equation*}%
This ratio is clearly measuring the amount by which the needlet coefficient
localized at $\left\{ \xi _{jk}\right\} $ is corrupted by the presence of
missing observations. Note that, due to the localization properties of spin
needlets, $s_{jk}$ will converge to unity for all pixels $\left\{ \xi
_{jk}\right\} $ which are outside the masked regions. We can now define the
thresholded parameters
\begin{equation*}
\beta _{jk;2}^{\ast }=\beta _{jk;2}\mathbb{I}(s_{jk}>t_{j})\text{ ,}
\end{equation*}%
where the function $\mathbb{I}(s_{jk}>t_{j})$ takes the value one if $%
s_{jk}>t_{j},$ zero otherwise; $t_{j}$ is a thresholding parameter which is
assumed to converge to one as $j\rightarrow \infty .$ We can then consider
the reconstructed maps%
\begin{equation}
\widetilde{E}^{\ast }(\hat{\gamma})\equiv -\frac{1}{2}\left\{ \sum_{jk}\left[
\beta _{jk;2}^{\ast }\varphi _{jk}(\hat{\gamma})+\overline{\beta
_{jk;2}^{\ast }}\overline{\varphi _{jk}}(\hat{\gamma})\right] \right\} \text{
,}  \label{recappe}
\end{equation}%
\begin{equation}
\widetilde{B}^{\ast }(\hat{\gamma})\equiv \frac{i}{2}\sum_{jk}\left[ \beta
_{jk;2}^{\ast }\varphi _{jk}(\hat{\gamma})-\overline{\beta _{jk;2}^{\ast }}%
\overline{\varphi _{jk}}(\hat{\gamma})\right] \text{ .}  \label{recappb}
\end{equation}%
In view of (\ref{betapp}), we expect $\widetilde{E}^{\ast}(\hat{\gamma})\simeq \widetilde{E}(%
\hat{\gamma}),$ $\widetilde{B}^{\ast}(\hat{\gamma})\simeq
\widetilde{B}(\hat{\gamma});$ the validity of these approximations
depends upon the distance of $\hat{\gamma}$ from the masked region
and the cosmic variance (i.e., the lower the value of $\left\{
C_{l}^{EE},C_{l}^{BB}\right\} $ at low multipoles, the better the
approximation). The accuracy of this approach will be tested
extensively in a future publication.

It should be noted that for the $\widetilde{E}$ and $\widetilde{B}$ fields
in (\ref{enorm1}, \ref{bnorm1}) we used the same definition and notation as
given by (\cite{wiaux06}); this definition for the scalar components seems
somewhat natural as it follows directly from the application of the spin
raising-spin lowering operation to polarization data. More commonly, in the
CMB literature the $E$ and $B$ random fields are defined instead as
\begin{equation*}
\widehat{E}(\hat{\gamma})=\sum_{lm}a_{lm;E}Y_{lm}(\hat{\gamma})\text{ , }%
\widehat{B}(\hat{\gamma})=\sum_{lm}a_{lm;B}Y_{lm}(\hat{\gamma})\text{ ;}
\end{equation*}%
i.e., these fields differ by a normalization factor $\sqrt{(l+2)!/(l-2)!}$
multiplying the random spherical harmonic coefficients $\left\{
a_{lm;E},a_{lm;B}\right\} $. It is indeed possible to obtain needlet
approximations of these maps, by focussing on
\begin{eqnarray*}
\widehat{E}^{\ast }(\hat{\gamma}) &\equiv &-\frac{1}{2}\left\{ \sum_{jk}%
\left[ \beta _{jk;2}^{\ast }\widehat{\varphi }_{jk}(\hat{\gamma})+\overline{%
\beta _{jk;2}^{\ast }}\overline{\widehat{\varphi }_{jk}}(\hat{\gamma})\right]
\right\} \text{ ,} \\
\widehat{B}^{\ast }(\hat{\gamma}) &\equiv &\frac{i}{2}\sum_{jk}\left[ \beta
_{jk;2}^{\ast }\widehat{\varphi }_{jk}(\hat{\gamma})-\overline{\beta
_{jk;2}^{\ast }}\overline{\widehat{\varphi }_{jk}}(\hat{\gamma})\right]
\text{ ,}
\end{eqnarray*}%
where%
\begin{equation}
\widehat{\varphi} _{jk}(\hat{\gamma})\overset{def}{=}\sqrt{\lambda _{jk}} \sum_{\ell m}b(\frac{\ell }{B^{j}}%
)\left\{ Y_{\ell m}(\hat{\gamma})\right\} \left\{ \overline{_{2}Y_{\ell m}}%
(\xi _{jk})\right\} \text{ .}  \label{finodif}
\end{equation}%
It should be noted however that $\widehat{E}(\hat{\gamma}),\widehat{B}(\hat{%
\gamma})$ do not follow trivially by application of the spin raising - spin
lowering operators to polarization data; likewise, and as opposed to (%
\ref{fidif}), (\ref{finodif}) is not the outcome of differentiation by $%
\partial $ ($\overline{\partial })$ on spin needlets, so that its
properties are less clear. (Because of (\ref{fidif1}), (\ref{fidif})
is well-localized by the results of \cite{spin-mat}.)

The results in the present subsection suggest that spin needlets may
allow for new statistical techniques when dealing with map
reconstruction, even besides those allowed by wavelet methods in the
scalar case. More generally, due to the poor localization properties
of standard spherical harmonics, in the
presence of masked regions and instrumental noise the construction of $E$ and $B$ from the coefficients $%
\left\{ \widetilde{a}_{lm;2}\right\} $ (see (\ref{maskfc})) may not
ensure satisfactory results (recall $\left\{
\widetilde{a}_{lm;2}\right\} $ need not be close to $\left\{
a_{lm;2}\right\} $ in any meaningful sense for partially observed
polarization maps). On the other hand,
in view of (\ref%
{betapp}) the coefficients $\left\{ \widetilde{%
\beta }_{jk}\right\} ,$ are much less affected by masked regions,
despite the fact that these quantities can themselves be viewed as
linear combinations of the spherical harmonic coefficients. This
suggests that spin needlet coefficients may be extremely useful when
attempting to build optimal polarization maps; for instance, it
seems natural to suggest the implementation of Internal Linear
Combination techniques based on wavelet coefficients (as done in the
scalar case by \cite{dela08}), without the need to go through scalar
maps as a first step. These developments are left for future work.

\subsection{Estimation of the angular power spectra $C_{l}^{EE},C_{l}^{BB}$}

We are now ready to address the estimation of the angular power spectra $%
C_{l}^{EE}$, $C_{l}^{BB}$ as separate quantities, which is clearly much more
meaningful from a physical point of view. For this purpose, it suffices to
write%
\begin{eqnarray*}
E_{j}(\hat{\gamma}) &=&-\frac{1}{2}\sum_{lm}b^{2}(\frac{l}{B^{j}})(a_{lm;2}+%
\overline{a}_{l-m;2})Y_{lm}(\hat{\gamma})\sqrt{\frac{(l+2)!}{(l-2)!}} \\
&=&\sum_{lm}b^{2}(\frac{l}{B^{j}})a_{lm;E}Y_{lm}(\hat{\gamma})\sqrt{\frac{%
(l+2)!}{(l-2)!}}\text{ ,}
\end{eqnarray*}%
and likewise%
\begin{equation*}
B_{j}(\hat{\gamma})=\sum_{lm}b^{2}(\frac{l}{B^{j}})a_{lm;B}Y_{\ell m}(\hat{%
\gamma})\sqrt{\frac{(l+2)!}{(l-2)!}}\text{ .}
\end{equation*}%
Clearly%
\begin{equation*}
<\int_{S^{2}}E_{j}^{2}(\hat{\gamma})d\widehat{\gamma }>=%
\int_{S^{2}}<E_{j}^{2}(\hat{\gamma})>d\widehat{\gamma }
\end{equation*}%
\begin{equation*}
=\sum_{l}b^{2}(\frac{%
l}{B^{j}})C_{l}^{EE}\frac{(l+2)!}{(l-2)!}\overset{def}{=}\widetilde{\Gamma }%
_{j}^{EE}\text{ ,}
\end{equation*}%
\begin{equation*}
<\int_{S^{2}}B_{j}^{2}(\hat{\gamma})d\widehat{\gamma }>=\sum_{l}b^{2}(\frac{l%
}{B^{j}})C_{l}^{BB}\frac{(l+2)!}{(l-2)!}=\widetilde{\Gamma }_{j}^{BB}\text{ .%
}
\end{equation*}%
In the presence of a mask, it is then natural to suggest the following
(unbiased) estimator:%
\begin{equation*}
\widetilde{C}_{j}^{EE}=\frac{\int_{S^{2}/M}(\widetilde{E}_{j}^{\ast })^{2}(%
\hat{\gamma})d\widehat{\gamma }}{\int_{S^{2}/M}d\widehat{\gamma }}\text{ , }%
\widetilde{C}_{j}^{BB}=\frac{\int_{S^{2}/M}(\widetilde{B}_{j}^{\ast }(\hat{%
\gamma}))^{2}d\widehat{\gamma }}{\int_{S^{2}/M}d\widehat{\gamma }}\text{ ;}
\end{equation*}%
here, the denominator is just a sky-fraction normalizing factor, whereas the
integrals can be easily approximated by sums in pixel spaces. It is obvious
that these estimators are unbiased for $\widetilde{\Gamma }_{j}^{EE}$%
,$\widetilde{\Gamma }_{j}^{BB};$ likewise, one might focus on
\begin{equation*}
\widehat{C}_{j}^{EE}=\frac{\int_{S^{2}/M}(\widehat{E}_{j}^{\ast }(\hat{\gamma%
}))^{2}d\widehat{\gamma }}{\int_{S^{2}/M}d\widehat{\gamma }}\text{ , }%
\widehat{C}_{j}^{BB}=\frac{\int_{S^{2}/M}(\widehat{B}_{j}^{\ast }(\hat{\gamma%
}))^{2}d\widehat{\gamma }}{\int_{S^{2}/M}d\widehat{\gamma }}\text{ ,}
\end{equation*}%
which are unbiased for the binned spectra $\sum_{l}b^{2}(l/B^{j})C_{l}^{EE},%
\sum_{l}b^{2}(l/B^{j})C_{l}^{BB}.$ A more complete set of properties and
implementation on polarization data will be provided elsewhere.

\section{A comparison with alternative constructions}

To compare our approach with possible alternative constructions (see
also \cite{cns}), let us consider again the scalar fields of
$\widehat{E}$- and $\widehat{B}$-modes, which we recall can be
written as%
\begin{equation*}
\widehat{E}(\hat{\gamma}):=\sum_{lm}a_{lm;E}Y_{lm}(\hat{\gamma})\text{ , }%
\widehat{B}(\hat{\gamma}):=\sum_{lm}a_{lm;B}Y_{lm}(\hat{\gamma})\text{ .}
\end{equation*}%
In the idealistic case of fully observed and noiseless maps for $F(\widehat{%
\gamma })=\left\{ Q(\hat{\gamma})+iU(\hat{\gamma})\right\} ,$ the
coefficients $\left\{ a_{lm;E},a_{lm;B}\right\} $ could be exactly recovered
and it would be natural to derive the scalar random needlet coefficients on $%
\widehat{E}(\hat{\gamma}),\widehat{B}(\hat{\gamma})$ as
\begin{equation}
\beta _{jk}^{E}=\sqrt{\lambda _{jk}}\sum_{lm}b\left( \frac{l}{B^{j}}\right)
a_{lm;E}Y_{lm}\left( \xi _{jk}\right) \text{ , }\beta _{jk}^{B}=\sqrt{%
\lambda _{jk}}\sum_{lm}b\left( \frac{l}{B^{j}}\right) a_{lm;B}Y_{lm}\left(
\xi _{jk}\right) \text{ .}  \label{bjke}
\end{equation}%
In other words, one might propose to define a wavelet construction by
deriving first the scalar maps, and then applying on them the standard
procedures. The purpose of this Section is to compare this approach with the
one we suggested earlier to show that they are not equivalent. More
precisely, we shall address the question about the localization properties
of (\ref{bjke}), and investigate the effect of the presence of a masked
region $M$.

Recall that for scalar fields, when $\xi _{jk}$ is ''far enough'' from $M$
we have the approximation%
\begin{eqnarray*}
\beta _{jk}^{E} &=&\sqrt{\lambda _{jk}}\sum_{lm}b\left( \frac{l}{B^{j}}%
\right) a_{lm;E}Y_{lm}\left( \xi _{jk}\right)  \\
&=&\int_{S^{2}}\widehat{E}(\hat{\gamma})\psi _{jk}(\hat{\gamma})d\hat{\gamma}%
\simeq \int_{S^{2}/M}\widehat{E}(\hat{\gamma})\psi _{jk}(\hat{\gamma})d\hat{%
\gamma}=\widetilde{\beta }_{jk}^{E}\text{ .}
\end{eqnarray*}%
In the presence of a masked region we obtain for instance%
\begin{eqnarray*}
\widetilde{\widetilde{\beta }}_{jk}^{E} &=&\sqrt{\lambda _{jk}}%
\sum_{lm}b\left( \frac{l}{B^{j}}\right) \widetilde{a}_{lm;E}Y_{lm}\left( \xi
_{jk}\right) \text{ ,} \\
\widetilde{a}_{lm}^{E} &=&-\frac{1}{2}\left\{ \int_{S^{2}/M}\left[ F(\hat{%
\gamma})\left( \overline{_{2}Y_{lm}}\left( \hat{\gamma}\right) \right) +\overline{F}(%
\widehat{\gamma })\left( \overline{_{-2}Y_{lm}}\left( \hat{\gamma}\right) \right) %
\right] d\hat{\gamma}\right\} \text{ .}
\end{eqnarray*}%
Now note that
\begin{equation*}
\widetilde{\widetilde{\beta }}_{jk}^{E}-\beta _{jk}^{E}=-\sqrt{\lambda _{jk}}%
\sum_{lm}b\left( \frac{l}{B^{j}}\right) \frac{1}{2}\left\{ \widetilde{a}%
_{lm}^{E}-a_{lm}^{E}\right\} Y_{lm}\left( \xi _{jk}\right)
\end{equation*}%
\begin{equation*}
=-\frac{1}{2}\sqrt{\lambda _{jk}}\sum_{lm}b\left(
\frac{l}{B^{j}}\right) \left\{ \int_{M}\left[ F(\hat{\gamma})\left(
\overline{_{2}Y_{lm}}\left( \hat{ \gamma} \right) \right)
+\overline{F}(\hat{\gamma})\left( \overline {_{-2}Y_{lm}}\left(
\hat{\gamma}\right) \right) \right] d\hat{\gamma}\right\}
Y_{lm}\left( \xi _{jk}\right)
\end{equation*}%
\begin{equation}
=-\frac{1}{2}\int_{M}\left[ F(\hat{\gamma})\overline {\chi} _{jk;2}\left( \hat{\gamma}%
\right) +\overline{F}(\hat{\gamma})\overline{\chi }_{jk;-2}\left( \hat{\gamma}%
\right) \right] d\hat{\gamma}\text{ ,}  \label{nosmall}
\end{equation}
where%
\begin{equation}
\chi _{jk;2}\left( \hat{\gamma}\right) =\sqrt{\lambda
_{jk}}\sum_{lm}b\left( \frac{l}{B^{j}}\right) \left(
_{2}Y_{lm}\left( \hat{\gamma}\right) \right) \overline{Y_{lm}}
\left( \xi _{jk}\right) \text{ ,}  \label{wrowav}
\end{equation}%

\begin{equation}
\chi _{jk;-2}\left( \hat{\gamma}\right) =\sqrt{\lambda
_{jk}}\sum_{lm}b\left( \frac{l}{B^{j}}\right) \left(
_{-2}Y_{lm}\left( \hat{\gamma}\right) \right)
\overline{Y_{lm}}\left( \xi _{jk}\right) \text{ .}  \label{wrowav2}
\end{equation}%

\begin{figure}
\begin{center}
\leavevmode
\epsfig {file=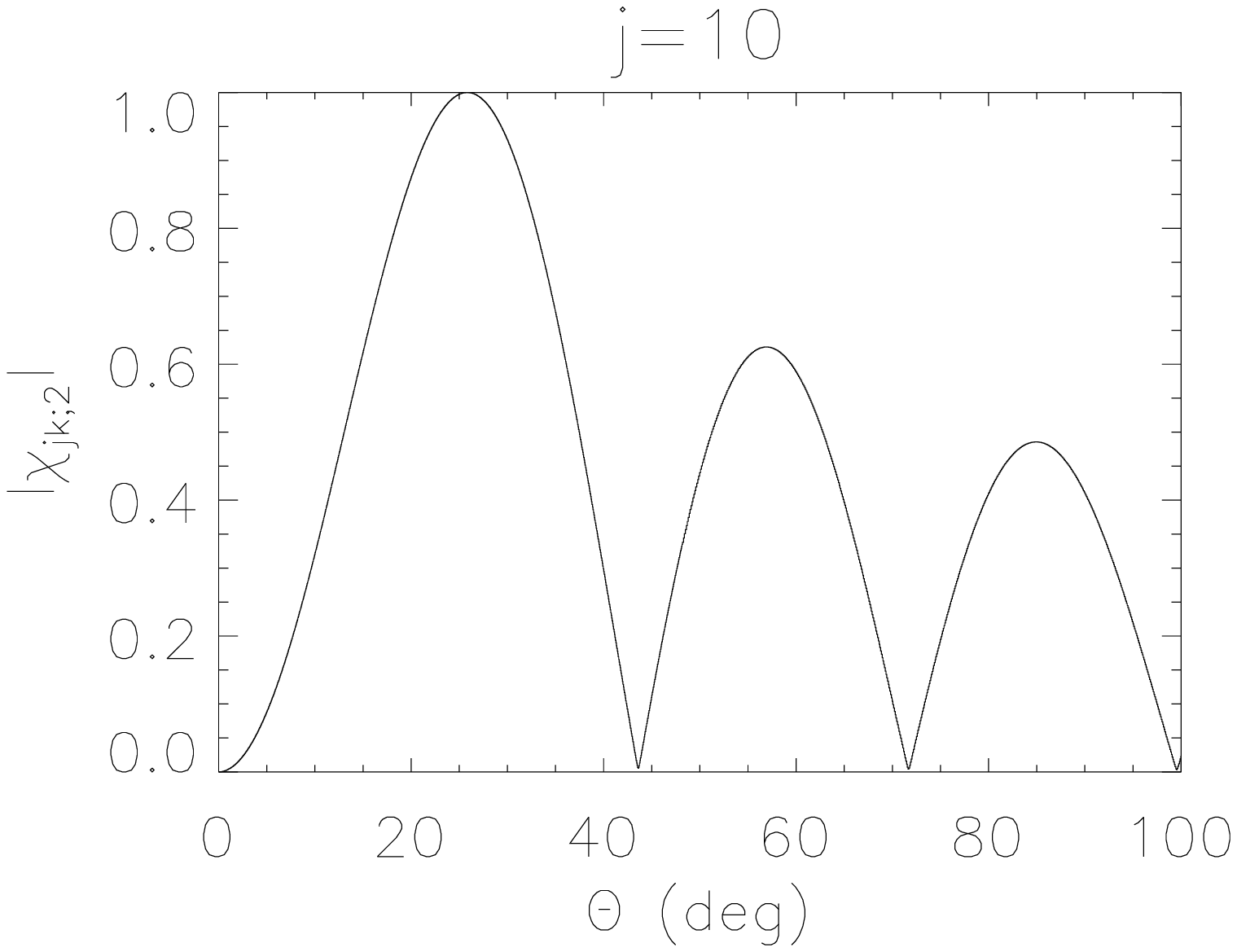,width=6cm,height=6cm}
\epsfig {file=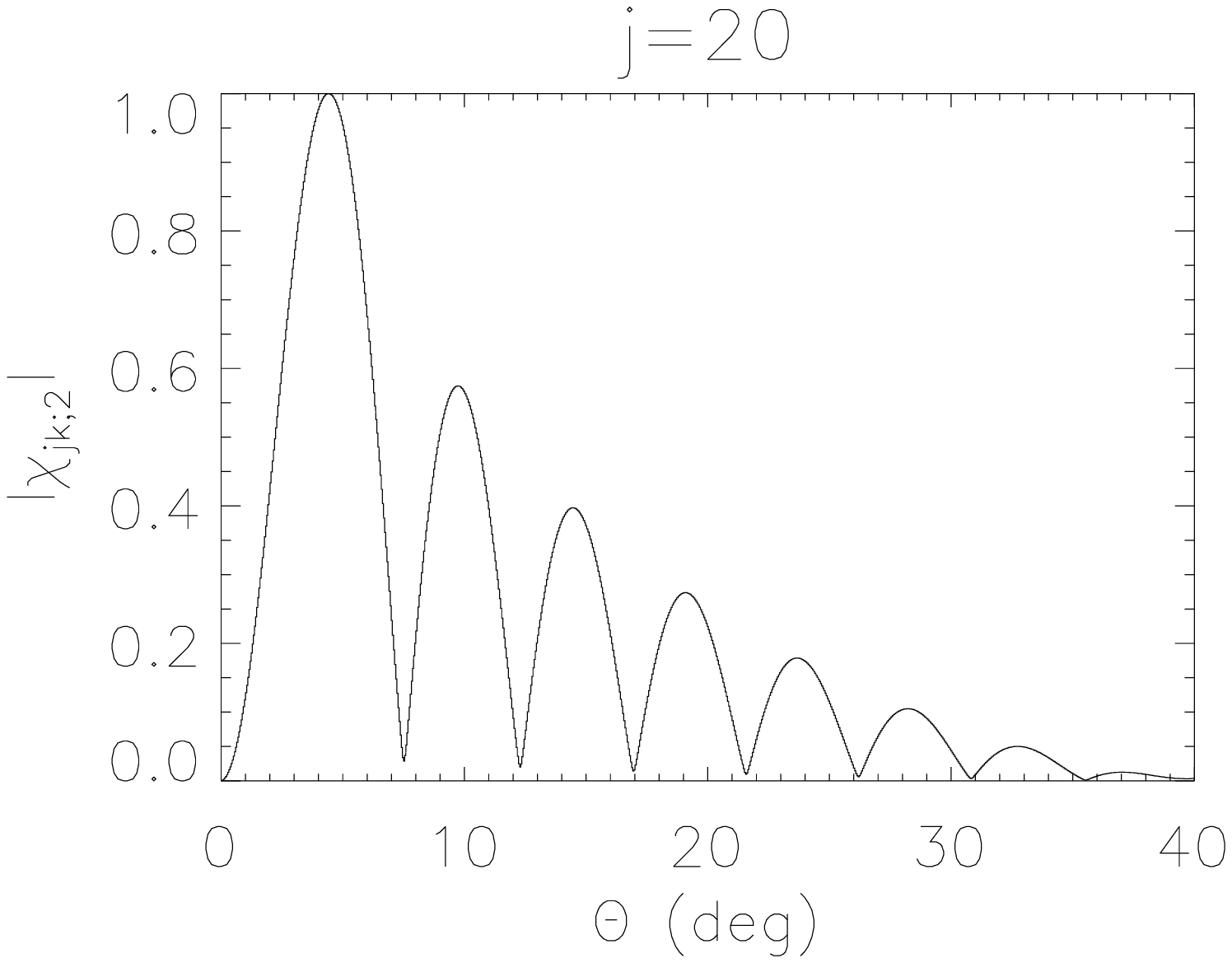,width=6cm,height=6cm}
\epsfig {file=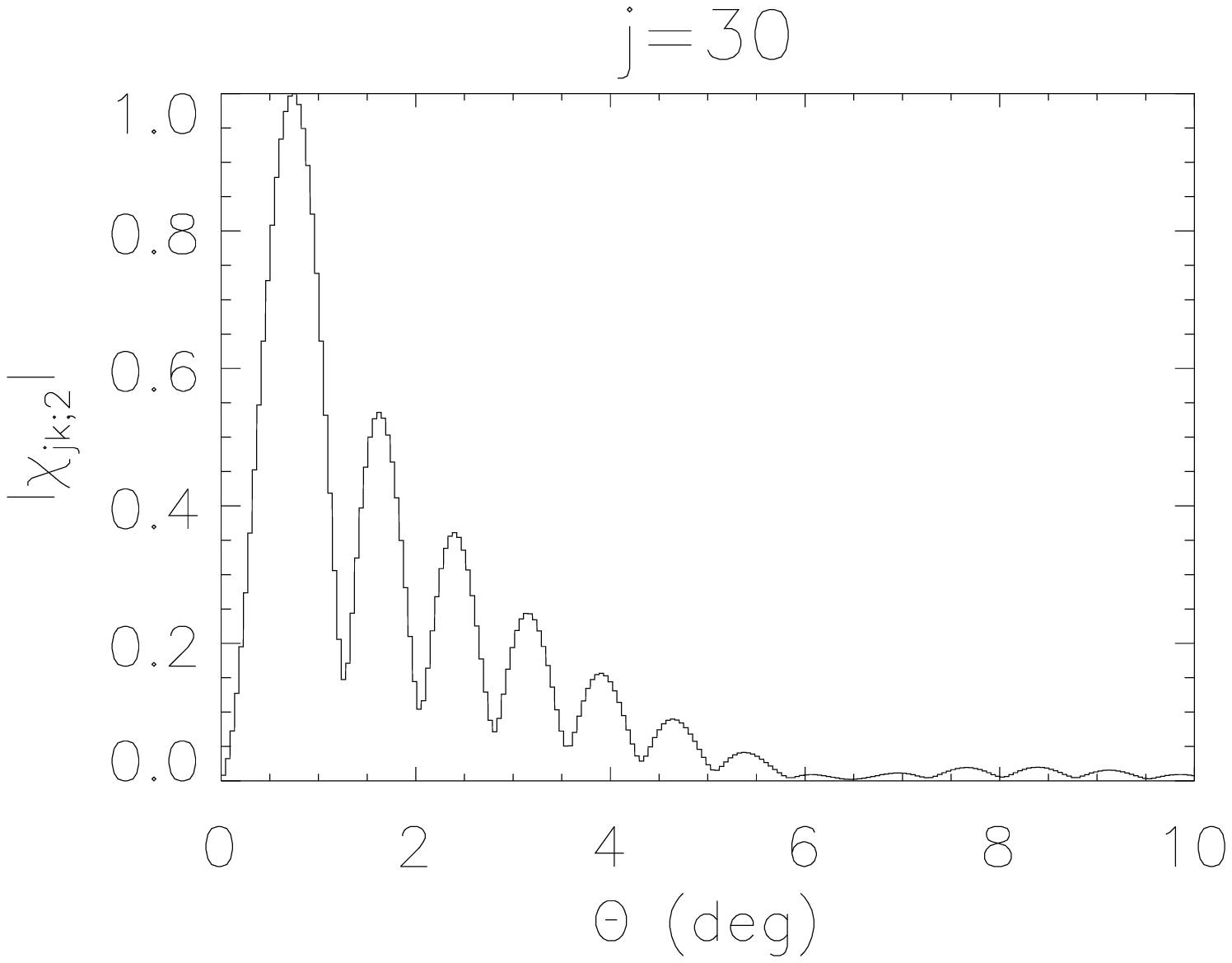,width=6cm,height=6cm}
\caption{The modulus of $\chi _{jk;2}(\hat{\gamma})$ for $B=1.2$ and $j=10$ (left plot), $j=20$ (right plot) and $j=30$ (lower plot). The angle $\theta$ is the distance from the center point.}
\label{fig:altneed}
\end{center}
\end{figure}

Now the properties of (\ref{wrowav}), (\ref{wrowav2}) are still
unknown and lack a rigorous mathematical exploration. Indeed the function $\sum_{m}\left( \overline{%
_{2}Y_{lm}\left( \hat{\gamma}\right) }\right) Y_{lm}\left( \xi
_{jk}\right) $ does not represent any form of projector on a proper
subspace, as it is the case for the corresponding expression $\sum_{m}\left( \overline{%
Y_{lm}\left( \hat{\gamma}\right) }\right) Y_{lm}\left( \xi
_{jk}\right) $ for the standard needlet construction in the scalar
case. In fact looking at figure \ref{fig:altneed} (compare with
figure \ref{fig:psi} for the spin needlets) we see that the shape of
$|\chi _{jk;2}\left( \hat{\gamma}\right)|$ is such that at the
origin the function has a zero, rather than a maximum, and then
increases away from the center. In other words, while spin needlets
enjoy a number of properties which are analogous to those in the
scalar case (good approximation properties in the presence of a mask
(see (\ref{betapp}) above), tight frame property, asymptotic
uncorrelation), these properties are not in general known to be
shared by constructing wavelets on the scalar random fields
$\widehat{E},\widehat{B}$ after implementing the spherical harmonic
transforms, in the presence of masked regions.

We believe these remarks provide a clear rationale for our
construction of spin needlets, which are directly embedded into the
geometric structure of polarization random fields.

\section*{Acknowledgements}
We are grateful to a referee for a careful reading of the manuscript
and useful suggestions, and to Xiaohong Lan for many useful
suggestions on a previous draft. FKH is also grateful for an OYI
grant from the Research Council of Norway.

\section{Appendix: Spin Spherical Harmonics}

In this Appendix, we review a few basic facts about the analysis
of polarization data. Our review will be quite informal, and we
will follow mainly the formalism by \cite{selzad}; we refer also
to \cite{kks96} for an
equivalent approach and a much more detailed treatment, see also \cite%
{wiaux06}. All the material discussed here is absolutely standard, and it is
reported just for completeness.

A standard route for the expansion of polarization random fields is based on
spin spherical harmonics (see for instance (\cite{selzad},\cite{wiaux06})).
Consider the standard representation of the group of rotations $SO(3)$ by
means of the Wigner's $D$-matrices, with elements%
\begin{equation*}
D_{mn}^{l}(\varphi ,\vartheta ,\chi )=e^{-im\varphi }d_{mn}^{l}(\vartheta
)e^{-in\chi },
\end{equation*}%
\begin{equation*}
d_{mn}^{l}(\vartheta )=\sum_{t=0\vee (m-n)}^{(l+m)\wedge (l-n)}\frac{(-1)^{t}%
\left[ (l+m)!(l-m)!(l+n)!(l-n)!\right] ^{1/2}}{(l+m-t)!(l-n-t)!t!(t+n-m)!}%
\left( \cos \frac{\vartheta }{2}\right) ^{2l+m-n-2t}\left( \sin \frac{%
\vartheta }{2}\right) ^{2t+n-m},
\end{equation*}%
where we recall that%
\begin{equation*}
Y_{lm}(\vartheta ,\varphi )\equiv \sqrt{\frac{2l+1}{4\pi }}%
D_{m0}^{l}(\varphi ,\vartheta ,0)\text{ .}
\end{equation*}%
The spherical harmonics of spin $n$ can then similarly be defined as
\begin{eqnarray}
_{n}Y_{lm}(\vartheta ,\varphi ) &\equiv &(-1)^{n}\sqrt{\frac{2l+1}{4\pi }}%
D_{m(-n)}^{l\ast }(\varphi ,\vartheta ,0)  \label{sad1} \\
&=&(-1)^{n}\sqrt{\frac{2l+1}{4\pi }}d_{m(-n)}^{l}(\vartheta )e^{im\varphi }.
\label{sad2}
\end{eqnarray}%
An important property we shall repeatedly use is

\begin{equation}
\sum_{m=-l}^{l}\left\{ _{n}Y_{l,m}(\vartheta ,\varphi )\right\} \left\{ _{n}%
\overline{Y_{l,m}(\vartheta ,\varphi )}\right\} =\frac{2l+1}{4\pi
}\text{ , }  \label{kerpro}
\end{equation}%
see for instance \cite{spin-mat}, Theorem 2.8.

Under a rotation around the tangent plane at the point $(\vartheta ,\varphi
) $, $_{n}Y_{lm}(\vartheta ,\varphi )$ transforms into $e^{-in\chi
}Y_{lm}(\vartheta ,\varphi )$, i.e. as a spin $n$ function. Spin $n$
functions can be defined iteratively by means of the spin raising and
lowering operators%
\begin{eqnarray}
\partial \left( _{n}G(\vartheta ,\varphi )\right) &:&=\left[ -\sin
^{n}\vartheta \left( \frac{\partial }{\partial \vartheta }+\frac{i}{\sin
\vartheta }\frac{\partial }{\partial \varphi }\right) \sin ^{-n}\vartheta %
\right] \left( _{n}G(\vartheta ,\varphi )\right) \text{ ,}  \label{spirai} \\
\overline{\partial }\left( _{n}G(\vartheta ,\varphi )\right) &:&=\left[
-\sin ^{-n}\vartheta \left( \frac{\partial }{\partial \vartheta }-\frac{i}{%
\sin \vartheta }\frac{\partial }{\partial \varphi }\right) \sin
^{n}\vartheta \right] \left( _{n}G(\vartheta ,\varphi )\right) \text{ .}
\label{spilow}
\end{eqnarray}%
We have for instance%
\begin{equation*}
\partial \left( _{n}Y_{lm}(\vartheta ,\varphi )\right) =\left[ (l-n)(l+n+1)%
\right] ^{1/2}\left( _{n+1}Y_{lm}(\vartheta ,\varphi )\right) \text{ ,}
\end{equation*}%
whence%
\begin{equation*}
\left( \partial \right) ^{2}\left( Y_{lm}(\vartheta ,\varphi )\right) =\left[
l(l+1)\right] ^{1/2}\left[ (l-1)(l+2)\right] ^{1/2}\left(
_{2}Y_{lm}(\vartheta ,\varphi )\right) \text{ ,}
\end{equation*}%
which generalizes iteratively to (for $0\leq n\leq l)$
\begin{equation*}
_{n}Y_{lm}(\vartheta ,\varphi )=\left[ \frac{(l-n)!}{(l+n)!}\right]
^{1/2}\left( \partial \right) ^{n}\left( Y_{lm}(\vartheta ,\varphi )\right)
\text{ ,}
\end{equation*}%
and (for $0\geq n\geq -l)$%
\begin{equation*}
_{n}Y_{lm}(\vartheta ,\varphi )=\left[ \frac{(l+n)!}{(l-n)!}\right]
^{1/2}(-1)^{n}\left( \overline{\partial }\right) ^{n}\left( Y_{lm}(\vartheta
,\varphi )\right) \text{ .}
\end{equation*}%
More explicitly%
\begin{equation*}
_{2}Y_{lm}(\vartheta ,\varphi )=\left[ \frac{(l-2)!}{(l+2)!}\right] ^{1/2}%
\left[ -\sin \vartheta \left( \frac{\partial }{\partial \vartheta }+\frac{i}{%
\sin \vartheta }\frac{\partial }{\partial \varphi }\right) \sin
^{-1}\vartheta \right] \left[ -\left( \frac{\partial }{\partial \vartheta }+%
\frac{i}{\sin \vartheta }\frac{\partial }{\partial \varphi }\right) \right]
Y_{lm}(\vartheta ,\varphi )
\end{equation*}%
\begin{equation*}
=\left[ \frac{(l-2)!}{(l+2)!}\right] ^{1/2}\left[ -\sin \vartheta \left(
\frac{\partial }{\partial \vartheta }+\frac{i}{\sin \vartheta }\frac{%
\partial }{\partial \varphi }\right) \sin ^{-1}\vartheta \right] \left[
\frac{m}{\sin \vartheta }Y_{lm}(\vartheta ,\varphi )-\frac{\partial }{%
\partial \vartheta }Y_{lm}(\vartheta ,\varphi )\right] \text{ .}
\end{equation*}%
Simple computations yield%
\begin{equation}
\sqrt{\frac{(l+2)!}{(l-2)!}}\left\{ _{\pm 2}Y_{lm}\right\} =\left[ \frac{%
m^{2}}{\sin ^{2}\vartheta }-\frac{2m}{\sin \vartheta }+2m\frac{\cot
\vartheta }{\sin \vartheta }-\cot \vartheta \frac{\partial }{\partial
\vartheta }+\frac{\partial ^{2}}{\partial \vartheta ^{2}}\right] Y_{lm}\text{
.}  \label{spinws}
\end{equation}%
We can then introduce the $E$ and $B$ scalar components, whose random
coefficients coefficients can be obtained (see (\cite{selzad}$,$\cite%
{wiaux06})) evaluating the \emph{spin 2} transforms
\begin{equation}
_{\pm 2}\widehat{Q}_{lm}:=\int_{S^{2}}Q(x)\left\{ _{\pm 2}Y_{lm}(x)\right\}
dx\text{ , }_{\pm 2}\widehat{U}_{lm}:=\int_{S^{2}}U(x)\left\{ _{\pm
2}Y_{lm}(x)\right\} dx\text{ ,}  \label{fourinv3}
\end{equation}%
and then proceeding to evaluate the coefficients%
\begin{eqnarray}
a_{lm}^{E} &=&-\frac{1}{2}\left[ \left( _{2}\widehat{Q}_{lm}+i_{2}\widehat{U}%
_{lm}\right) +\left( _{-2}\widehat{Q}_{lm}-i_{-2}\widehat{U}_{lm}\right) %
\right] \text{ ,}  \label{fourinv4} \\
a_{lm}^{B} &=&\frac{i}{2}\left[ \left( _{2}\widehat{Q}_{lm}+i_{2}\widehat{U}%
_{lm}\right) -\left( _{-2}\widehat{Q}_{lm}-i_{-2}\widehat{U}_{lm}\right) %
\right] \text{ .}  \label{fourinv5}
\end{eqnarray}

\end{document}